\documentclass[rmp,noshowpacs]{revtex4}
\usepackage{amsfonts,amssymb,latexsym,amsmath,amscd}
\usepackage{times,mathptm}
\usepackage[ps2pdf,colorlinks]{hyperref}
\usepackage{anysize}
\marginsize{1in}{1in}{1in}{1in}
\newcommand{\ket}[1]{|#1\rangle}
\newcommand{\bra}[1]{\langle#1|}

\newenvironment{example}
    {
    \smallskip
    \refstepcounter{theorem}
    \noindent
    {\bf Example \Roman{section}.\arabic{theorem}} \ \ }
    {\hspace*{\fill}{$\Diamond$}
    \smallskip}

\newenvironment{remark}
    {
    \smallskip
    \refstepcounter{theorem}
    \noindent
    {\bf Remark \Roman{section}.\arabic{theorem}} \ \ }
    {\hspace*{\fill}{$\Diamond$}
    \smallskip}

\newenvironment{definition}
    {
    \smallskip
    \refstepcounter{theorem}
    \noindent
    {\bf Definition \Roman{section}.\arabic{theorem}} \ \ }
    {\hspace*{\fill}{\ }
    \smallskip}

\newenvironment{appendix_definition}
    {
    \smallskip
    \refstepcounter{theorem}
    \noindent
    {\bf Definition \Alph{section}.\arabic{theorem}} \ \ }
    {\hspace*{\fill}{\ }
    \smallskip}

\newenvironment{scholium}
    {
    \smallskip
    \refstepcounter{theorem}
    \noindent
    {\bf Scholium \Roman{section}.\arabic{theorem}} \it \ \ }
    {\hspace*{\fill}{\ }
    \smallskip}

\newenvironment{proof}[1][]
    {
    \noindent
    {\bf Proof{#1}:  }
    }
    {\hspace*{\fill}{$\Box$}\smallskip}

\newenvironment{sketch}[1][]
    {
    \noindent
    {\bf Sketch{#1}:  }
    }
    {\hspace*{\fill}{$\Box$}\smallskip}

    {
    \noindent
    {\bf Reference:  }
    }
    {\hspace*{\fill}{$\odot$}\smallskip}

\newtheorem{theorem}{Theorem}[section]
\newtheorem{proposition}[theorem]{Proposition}
\newtheorem{lemma}[theorem]{Lemma}
\newtheorem{corollary}[theorem]{Corollary}

\begin{document}
\title{Canonical Decompositions of $n$-qubit Quantum Computations and 
Concurrence}

\author{Stephen S. Bullock}
\address{Mathematical and Computer Sciences Division, National Institute of Standards and 
Technology, Gaithersburg, MD,  20089 } \email{\tt stephen.bullock@nist.gov}
\author{Gavin K. Brennen}
\address{Quantum Processes Group, National Institute of Standards and Technology, Gaithersburg, MD, 20089} 
\email{\tt gavin.brennen@nist.gov}

\begin{abstract}
The two-qubit canonical decomposition $SU(4) = [SU(2) \otimes SU(2)]
\Delta [SU(2) \otimes SU(2)]$ writes any two-qubit unitary operator
as a composition of a local unitary, a relative phasing of
Bell states, and a second local unitary.  Using Lie theory,
we generalize this to an $n$-qubit decomposition,
the concurrence canonical decomposition ({\tt CCD}) 
$SU(2^n)=KAK$.  The group $K$ fixes a bilinear form related to the
concurrence, and in particular any unitary in $K$ preserves the
tangle $| \overline{\bra{\phi}}  (-i\sigma^y_1)\cdots (-i\sigma^y_n)
 \ket{\phi} |^2$ for $n$ even.  Thus, the {\tt CCD} 
shows that any $n$-qubit unitary is a composition of a unitary operator
preserving this $n$-tangle, a unitary operator in $A$ which applies relative
phases to a set of {\tt GHZ} states, and a second unitary operator which 
preserves the tangle.

As an application, we study the extent to which a large, random unitary
may change concurrence.
The result states that for a randomly chosen $a \in A \subset SU(2^{2p})$,
the probability that $a$ carries a state of tangle $0$ to a state
of maximum tangle approaches $1$ as the even number of qubits approaches
infinity.  Any $v=k_1 a k_2$ for such an $a\in A$
has the same property.  Finally, although
$| \overline{\bra{\phi}}  (-i\sigma^y_1)\cdots (-i\sigma^y_n)
 \ket{\phi} |^2$ vanishes identically when the number of qubits is odd,
we show that a more complicated {\tt CCD} still exists in which $K$ is a
symplectic group.
\end{abstract}
\maketitle
\section{Introduction}

Entanglement is a unique feature of quantum systems that plays a key role in 
quantum information proccessing.
Much effort has gone into describing the entanglement 
present in the state of a quantum system composed of two or more 
measurably distinct subsystems.  Because different amounts of entanglement 
may be shared among the various partitions of the
tensor factors of the Hilbert state space, there is 
no single measure of entanglement that captures all non-local 
correlations for many-particle systems.  Rather, the number of partitions
of the tensor factors grows exponentially with the number of factors
themselves.  Thus, it is reasonable to guess that same is true for the
number of useful entanglement measures.  In fact, the situation is yet
more complicated.  Many reasonable definitions create
uncountably many entanglement types, which thus
may not be associated to countable collections of partitions or monotones.

Nevertheless, it is interesting to consider how much
entanglement is created by a given unitary evolution $U$ of an
$n$-qubit state space.  To achieve this in a limited context, we focus on 
a single multi-qubit entanglement measure, the $n$-concurrence 
\cite{Wong:01}.  Using Lie theory, we may decompose a unitary 
operator acting on $n$ qubits into a form such that the entangling power 
of the unitary with respect to this measure is manifest.

The $n$-tangle and its square root, the $n$-concurrence, are two of several proposed
multiqubit entanglement measures.  Others include polynomial invariants 
which involve moments of the reduced state eigenvalues \cite{Barnum:01}, 
the Schmidt measure \cite{Eisert:01} which is related to the minimum number 
of terms in the product state expansion of a state, the $Q$ measure 
\cite{Meyer:02} which is related to the average purity each qubit's
reduced state, and {\bf GAVIN ADDS SOMETHING.}  \cite{EggelingWerner:01}
A further measure makes use of hyperdeterminants \cite{Miyake:02};
this powerful technique makes computation 
difficult in more than six qubits.
The \emph{concurrence} $C_n$ is originally introduced in the two-qubit
case \cite{Hill:97}.  It is generalized to a measure on two systems of 
arbitrarily many
dimensions in \cite{Rungta:01} and extends to $n$-qubits 
\cite{Wong:01}.  

We now consider the quantitative expression for the concurrence.
Suppose a quantum state space of data for a quantum computer.
Specifically, fix $n$ as the number of qubits, $N=2^n$.  Throughout,
we use $\ket{j}$ not to denote the state of a qudit but rather as
an abbreviated multi-qubit state via binary form.  For example, in three 
qubits $\ket{5}=\ket{101}=\ket{1}\otimes \ket{0}\otimes \ket{1}$.
We write $\mathcal{H}_n = \mbox{span}_{\mathbb{C}}\{ \ket{j} \; ; \;
0 \leq j \leq N-1\}$ for the $n$-qubit Hilbert state space.
Then the concurrence is a map
\hbox{$C_n:\mathcal{H}_n \rightarrow [0,\infty)$}
given by
$C_n(\ket{\psi}) = | \overline{\bra{\psi}} (-i\sigma^y_1)\cdots(-i\sigma^y_n)
\ket{\psi}|$.  Note that the expression inside the complex norm is in
general not real.
A related entanglement measure $\tau_n = C_n^2$ is known as the
\emph{$n$-tangle} when $n$ is even \footnote{It is not clear how to
recover the celebrated $3$-tangle from this construction.  By 
this definition we will see that $\tau_n \equiv 0$ for all odd $n$,
so references to $\tau_n$ in the present paper will suppose that is not
the case.}.
For $\bra{\psi} \psi \rangle = 1$,
the $n$-tangle of a state $\ket{\psi}$ assumes real values in the range 
$0\leq \tau_{n} \leq1$.
It is moreover an entanglement monotone \cite{Vidal:00}, as any good measure 
should be.
This means in particular that 
$\tau_{n}:\mathcal{H}_n \rightarrow [0,\infty)$ vanishes on full tensor
products of local states, and moreover that
$\tau_n(  v_1 \otimes v_2 \otimes \cdots \otimes v_n \ket{\psi})=
\tau_n(\ket{\psi})$
for any $v_1 \otimes v_2 \otimes \cdots \otimes v_n \in \otimes_1^n SU(2)$.  
We show in Appendix \ref{app:monotone} that the 
$n$-concurrence is also an entanglement monotone.

The $n$-concurrence only detects certain kinds of entanglement.  
Specifically, while it 
returns zero on all separable states, it may also return zero on certain 
non-separable states.  We illustrate the monotone's behavior by example.
First, the $n$-partite Greenberger-Horne-Zeilinger ({\tt GHZ}) state
$|GHZ_{n}\rangle=(1/\sqrt{2})(|0_{1}\ldots 0_{n}\rangle+|1_{1}\ldots 
1_{n}\rangle)$ has maximal $n$-concurrence while 
$C_{n}(|GHZ_{n-1}\rangle\otimes|0\rangle_n)=0$.  As a second example,
the  generalized $\ket{W}$ state given by
$\ket{W}=(1/\sqrt{n})(\ket{10\ldots0}+\ket{010\ldots0}+\cdots+
\ket{0\ldots01})$ 
has zero $n$-concurrence despite being entangled.  States 
with subglobal entanglement can also 
assume maximal $n$-concurrence; 
$C_{n}(|GHZ_{n}\rangle)=
C_{n}(|GHZ_{n/2}\rangle\otimes|GHZ_{n/2}\rangle)=1$.  Generally, the 
$n$-concurrence seeks out superpositions between
a state and its binary bit flip.   

We extend these definitions by introducing a complex bilinear form,
the \emph{concurrence form} \hbox{$\mathcal{C}_n: \mathcal{H}_n \times
\mathcal{H}_n \rightarrow \mathbb{C}$}.  Here, complex bilinear means
the function is linear when restricted to each variable.  
The antisymmetric concurrence form
$\mathcal{C}_n(-,-)$ is nonzero even in the case $n$ is odd, although
of course $C_{2p-1} \equiv 0$ since $\mathcal{C}_{2p-1}(\ket{\psi},\ket{\psi})=
-\mathcal{C}_{2p-1}(\ket{\psi},\ket{\psi})=0$.

\begin{definition}
\hspace{-2mm}
The concurrence form $\mathcal{C}_n : \mathcal{H}_n \times \mathcal{H}_n
\rightarrow \mathbb{C}$ is given by
$\mathcal{C}_n(\ket{\psi},\ket{\phi}) = \overline{\bra{\psi}}
(-i\sigma^y_1)(-i\sigma^y_2)\cdots(-i\sigma^y_n) \ket{\phi}$.
Note the complex conjugation of the lead bra is required for complex
linearity (rather than antilinearity) in the first variable.
The concurrence quadratic form is
$Q_n^C(\ket{\psi})=\mathcal{C}_n(\ket{\psi},\ket{\psi})$,
so that $C_n(\ket{\psi})=|Q_n^C(\ket{\psi})| = \sqrt{\tau_n({\ket{\psi}})}$.
Note that $Q_n^C$ is a complex quadratic polynomial on the vector space
$\mathcal{H}_n$.
\end{definition}

The main technique of this paper is to build a new matrix decomposition of
the Lie group of global phase normed quantum computations $SU(N)$.
It is optimized for the study of the concurrence and $n$-tangle and
generalizes the two-qubit canonical decomposition
\cite{Makhlin:00,KrausCirac:00,Khaneja:01,Lewenstein:01,Bullock:03,Zhang:03}
\begin{equation}
SU(4)=[SU(2)\otimes SU(2)] \Delta [SU(2)\otimes SU(2)]
\end{equation}
Here, the commutative group $\Delta$ applies 
relative phases to a ``magic basis'' \cite{Bennett:96, Hill:97,
Khaneja:01,Lewenstein:01,Bullock:03}
of phase-shifted Bell states.  This two-qubit canonical decomposition
is used to the study of the entanglement capacity of
two-qubit operations \cite{Zhang:03}, to
building efficient (small) circuits in two qubits
\cite{Bullock:03, Vidal:03, Shende:03},
and to classify which two-qubit unitary operators require fewer than
average multiqubit interactions \cite{Vidal:03,Shende:03}.

The canonical decomposition is itself an example of the $G=KAK$
metadecomposition theorem of Lie theory 
\cite[thm8.6,\S VII.8]{Helgason:01}.
This theorem produces a decomposition of an input semisimple
Lie group $G$ given two further inputs:
\begin{itemize}
\item  a Cartan involution
\cite[\S X.6.3,pg.518]{Helgason:01}
\hbox{$\theta:\mathfrak{g}\rightarrow \mathfrak{g}$} for
$\mathfrak{g}=\mbox{Lie}(G)$.  By definition, $\theta$ satisfies
(i) $\theta^2 = {\bf 1}$ and (ii)
$\theta[X,Y]=[\theta X, \theta Y]$ for all $X,Y \in \mathfrak{g}$.
We write $\mathfrak{g}=\mathfrak{p} \oplus \mathfrak{k}$ for the
decomposition of $\mathfrak{g}$ into the $+1$ and $-1$ eigenspace
of $\theta$.
\item a commutative subalgebra $\mathfrak{a} \subset \mathfrak{p}$ which
is maximal commutative in $\mathfrak{p}$.
\end{itemize}
Then write $K = \mbox{exp } \mathfrak{k}$, $A = \mbox{exp } \mathfrak{a}$,
where for linear $G \subset GL(n,\mathbb{C})$ the exponential coincides
with the matrix power series on each of the Lie subalgebras
$\mathfrak{k}$, $\mathfrak{a}$.  The theorem asserts then that
$G = KAK = \{ k_1 a k_2 \; ; \; k_1,k_2 \in K, a \in A \}$.

For example, the cononical decomposition of $SU(4)$ arises as follows.
Take \hbox{$\theta: \mathfrak{su}(4) \rightarrow \mathfrak{su}(4)$}
by $\theta(X)=(-i \sigma_1^y)(-i\sigma^y_2) \bar{X}
(-i \sigma_1^y)(-i\sigma^y_2)$ and
\begin{equation}
\mathfrak{a} = \mbox{span}_{\mathbb{R}} 
\bigg\{ 
\ \ i \ket{0}\bra{0}-i\ket{1}\bra{1}-i\ket{2}\bra{2}+i\ket{3}\bra{3}, \ 
i\ket{0}\bra{3}+i\ket{3}\bra{0},\ i\ket{1}\bra{2}+i\ket{2}\bra{1} 
\ \ \bigg\}
\end{equation}
We extend this particular construction to $n$-qubits.

\vbox{
\begin{definition}
\label{def:KAK}
Let $S= (-i\sigma^y_1)(-i\sigma^y_2)\cdots(- i\sigma^y_n)$.
Define $\theta:\mathfrak{su}(2^n) \rightarrow \mathfrak{su}(2^n)$
by $\theta(X)=S^{-1} \bar{X} S = (-1)^n S\bar{X}S$.  Then
$\mathfrak{k}$ denotes the $+1$-eigenspace of $\theta$ while
$\mathfrak{p}$ denotes the $-1$-eigenspace.  Finally, in case
$n$ is even we define
\begin{equation}
\begin{array}{lcl}
\mathfrak{a} & = &
\mbox{span}_{\mathbb{R}}
\big( \{ \ 
i \ket{j}\bra{j}+i \ket{N-j-1}\bra{N-j-1} -i
\ket{j+1}\bra{j+1}-i\ket{N-j-2}\bra{N-j-2} \ ; \ 
0 \leq j \leq 2^{n-1}-2 \ 
\} \\
& & \quad \quad \quad \quad \sqcup \quad
\{ \ i \ket{j}\bra{N-j-1} + i\ket{N-j-1} \bra{j}  \ ; \
0 \leq j \leq 2^{n-1}-1 \ \} , \big) \mbox{  in case $n$ even}
\end{array}
\end{equation}
with $A=\mbox{exp}\; \mathfrak{a}$.  In case $n$ odd, we drop the
second set:
  \begin{equation}
\begin{array}{lcl}
\mathfrak{a} & = &
\mbox{span}_{\mathbb{R}}
\big( \{ \ 
i \ket{j}\bra{j}+i \ket{N-j-1}\bra{N-j-1} -i
\ket{j+1}\bra{j+1}-i\ket{N-j-2}\bra{N-j-2} \ ; \ 
0 \leq j \leq 2^{n-1}-2 \ 
\} \big) \\
& & \mbox{      in case $n$ odd} \\
\end{array}
\end{equation}
Modulo checks reserved for the body, the \emph{concurrence
canonical decomposition ({\tt CCD})} in $n$-qubits is the resulting
matrix decomposition $SU(2^n)=KAK$.  Note that $n$ may be even or odd.
\end{definition}
}

In $n$-qubits, it is certainly \emph{not} the case that $K$ is the
Lie group of local unitaries.  Nonetheless, we prove momentarily by
direct computation that the local unitary group
$SU(2) \otimes SU(2) \otimes \cdots \otimes SU(2) \subset K$,
with strict containment in $n\geq 3$ qubits by a dimension count.
Moreover, for $n=2p$ an even number of qubits the concurrence canonical
decomposition is computable via an algorithm familiar from the
two-qubit case \cite{Bullock:03}, (see Appendix \ref{app:compute}).
The following theorem provides the key to interpreting this extended
canonical decomposition.

\vbox{
\begin{theorem}  
\label{thm:K_sym}
Let $K =\mbox{exp } \mathfrak{k}$ for $\mathfrak{k}$ the
$+1$-eigenspace of the Cartan involution $\theta(X)=S^{-1}\bar{X} S$.
{\em Then $K$ is the symmetry group of the concurrence 
form $\mathcal{C}_n$.}
Specifically, for $u \in SU(N)$,
\begin{equation}
(u \in K) \Longleftrightarrow
\big[ \ \ \mathcal{C}_n(u \ket{\phi}, u\ket{\psi}) =
\mathcal{C}_n(\ket{\phi}, \ket{\psi}) \ \quad  \mbox{for every }
\ket{\phi}, \ket{\psi} \in \mathcal{H}_n \big]
\end{equation}
Moreover, for $n$ even the concurrence form 
is symmetric.  In the even case, it restricts to
the usual dot product on the $\mathbb{R}$-span of a collection of
$n$ concurrence one states, and this $\mathbb{R}$ subspace of
$\mathcal{H}_n$ is preserved by $K$.  On the other hand, for $n$ odd
$\mathcal{C}_n$ is antisymmetric, i.e. a two-form.  Thus,
\begin{itemize}
\item $K \cong Sp(N/2)$, if $n$ is an odd number of qubits
\item $K \cong SO(N)$, if $n$ is an even number of qubits
\end{itemize}
\end{theorem}
}

\begin{remark}
Bremner et al \cite[thm5]{BremnerEtAl:03}
observe symplectic Lie algebras independently in a context
related to the above.  We explore this in more detail in a sequel
manuscript.
\end{remark}

This interpretation allows for an extension of prior work on the
entangling capacity of two-qubit unitaries \cite{Zhang:03}.
Here is the precise result:

\begin{definition}
\label{def:kappa}
The \emph{concurrence capacity} of a given $n$-qubit unitary operator
$v \in SU(N)$ is defined by
$\kappa(v)=\mbox{max} \{
C_n(v\ket{\psi}) \; ; \; {C_n(\ket{\psi})=0, \bra{\psi} \psi \rangle=1} \}$.
\end{definition}

\begin{corollary}[ of \ref{thm:K_sym}]
\label{cor:all_in_a}
Let $u=k_1 a k_2$ be the $n$-qubit canonical decomposition of
$u \in SU(N)$.  Then $\kappa(u)=\kappa(a)$.
\end{corollary}

Given the {\tt CCD}, 
the function $\kappa$ is properly viewed
as a function on the $A$ factor rather than on the entire group
of phase-normalized unitaries $SU(N)$.  Finally, a careful analysis of
$\kappa(a)$ for randomly chosen $a$ in $A$ produces the following, perhaps
surprising result.

\begin{theorem}
\label{thm:almost_all}
Suppose the number of qubits is even, i.e. $n=2p$.  
Then for large $p$
almost all $a \in A$ have maximal concurrence capacity.
Specifically, suppose we choose $a \in A$ at random per the probability
density function given by the unit normalized Haar measure $da$.  Then
\begin{equation}
\mbox{lim}_{p \rightarrow \infty} \mbox{Probability} [\kappa(a)=1] \ \ = \ \ 
\mbox{lim}_{p \rightarrow \infty}\  da({\{a \in A \; ; \; \kappa(a) = 1\}}) \ \ = \ \ 1
\end{equation}
\end{theorem}

We rephrase this result colloquially.  Suppose we think of those states
$\ket{\psi}$ in even qubits with $\tau_n(\ket{\psi}) =1$ as {\tt GHZ}-like.
Then as the even number of qubits grows large, almost every 
unitary evolution will be able to produce such 
a maximally concurrent {\tt GHZ}-like state
from some input state of $0$ concurrence.

\subsection*{Notation and Contents}

We provide some samples of our notation for the reader's convenience.
Throughout, $n$ is a number of qubits and $N=2^n$.
For $v = \sum_{j,k=0}^{N-1} v_{j,k}\ket{k}\bra{j}$,
we have the adjoint
$v^\dagger = \sum_{j,k=0}^{N-1} \bar{v}_{k,j} \ket{k}\bra{j}$.
We also require the transpose operation, most easily visualized in
matrix form as $(v_{j,k})^T = (v_{k,j})$.  Equivalently,
$v^T = \sum_{j,k=0}^{N-1} v_{j,k}\ket{j}\bra{k}$.  Thus
$v^\dagger=\bar{v}^T$.
Recall also the convention of collapsing the binary for an integer inside
the ket of a computational basis state.  We use lower rather
than upper case letters for most operators to avoid confusing them with
Lie groups denoted by capital letters.  The older term \emph{scholium}
is used to refer to a corollary of the proof of a theorem or proposition
rather than its formal statement.
Besides these conventions,
we follow the notations of either \cite{Nielsen:00} or
\cite{Helgason:01}.

The paper is structured as follows.  In \S \ref{sec:ccd}, we verify that
the conventions are Definition \ref{def:KAK} are appropriate for
invoking the $G=KAK$ theorem.  Having verified that the matrix decomposition
exists, \S \ref{sec:ccd} further describes \emph{entanglers} and
\emph{finaglers}, loosely similarity matrices which rotate the
{\tt CCD} onto more standard $KAK$ decompositions of $SU(N)$.  In
\S \ref{sec:capacity}, we discuss the concurrence
capacity and prove the properties of this capacity
asserted above.  The three appendices consecutively
(i) provide an algorithm for computing the {\tt CCD} given a matrix
$v \in SU(N)$, exclusively in the case $n$ is even,
(ii) argue that any two normalized states $\ket{\phi}$, $\ket{\psi}$ with
identical concurrence must have $k \ket{\phi}=\ket{\psi}$ for some
$k$ in the symmetry group $K$, and (iii) prove that
the concurrence $C_n(-)$ is an entanglement monotone.

\section{Entanglers, finaglers, and Cartan involutions of $\mathfrak{su}(N)$}
\label{sec:ccd}

This section has two goals.  First, we show our $KAK$ decomposition is
well-defined, by noting that $\theta$ is a Cartan involution,
checking by direct computation that $\mathfrak{a}$ is abelian,
and arguing that $\mathfrak{a}$ is maximal commutative.
Second, we prove Theorem \ref{thm:K_sym}.
There are generally two approaches to the theorem.  We could
recall standard Cartan involutions and $KAK$
decompositions from the literature.  
We will shortly construct similarity matrices $E_0$ and $F_0$ which
rotate the standard $G=KAK$ decompositions of $SU(N)$ onto the
{\tt CCD}, and we could simply appeal to these matrices and the
standard structures.  Alternately, (many) intrinsic computations
would suffice to check the required properties for $G=KAK$.  The present
approach is a compromise.  The argument
that the {\tt CCD} $SU(N)=KAK$ is well-defined 
is intrinsic, except for a single appeal to classification.  On the 
other hand, the classification of the 
$K$ groups uses similarity matrices.  As such, it 
is ultimately a change of basis in the $n$-qubit state space $\mathcal{H}_n$.

\subsection*{Properties of the {\tt CCD} $SU(N)=KAK$}

The following proposition is not used in the sequel.  However, we include
a direct proof due to its importance in guiding the
choice of $\theta$.  It simplifies an older argument and arose from
correspondence with P.Zanardi.

\begin{proposition}  
\label{prop:localsinK}
Let $K$ be as in Definition \ref{def:KAK}.  Then there is an inclusion
$SU(2) \otimes SU(2) \otimes \cdots \otimes SU(2)  \subset K$.

\end{proposition}

\begin{proof}
Recall
$i\sigma^x = i \ket{0} \bra{1} + i \ket{1}\bra{0}$,
$i\sigma^y = \ket{0} \bra{1} - \ket{1} \bra{0}$, and
$i\sigma^z = i \ket{0} \bra{0} - i\ket{1} \bra{1}$ forms a basis of
$\mathfrak{su}(2^1)$.  For the statement of the proposition,
it suffices to check $
\mbox{Lie} [\otimes_1^n SU(2)] = 
\mbox{span} \{ i \sigma^x_j, i \sigma^y_j, 
i \sigma^z_j \; ; \; 1 \leq j \leq n \}
\subset \mathfrak{k}$.  
We further recall the last item of Lemma \ref{lem:S_props}, as
well as the fact that the complex conjugates
of the Pauli matrices are
$\overline{i \sigma^x} = -i \sigma^x$,
$\overline{i \sigma^y} = i \sigma^y$, and
$\overline{i \sigma^z} = -i \sigma^z$.  
Then we wish to show that $\theta$ fixes every
$\sigma^x_j$, $\sigma^y_j$, and $\sigma^z_j$.  For this,
\begin{displaymath}
\begin{array}{lclclclcl}
(-1)^n \; S \; \overline{(i \sigma^x_j)} \; S & = & (-1)^n 
S \; (-i \sigma^x_j) \; S
& = & (-1)^n S^2 \; (i \sigma^x_j) & = & (i \sigma^x_j) \\
(-1)^n \; S \; \overline{(i \sigma^y_j)} \; S 
& = & (-1)^n S \; (i \sigma^y_j) \; S & = &
(-1)^n S^2 \; (i \sigma^y_j) & = & (i \sigma^y_j) \\
(-1)^n \; S \; \overline{(i \sigma^z_j)} \; S & = & 
(-1)^n S \; (-i \sigma^z_j) \; S & = &
(-1)^n S^2 (i \sigma^z_j) & = & (i \sigma^z_j) \\
\end{array}
\end{displaymath}
Hence each such infinitesimal unitary is in the $+1$
eigenspace of $\theta$.  This concludes the proof.
\end{proof}

We next note that $\theta$ is a Cartan involution.  Indeed,
direct computation shows that $\theta^2 = {\bf 1}$.  Moreover,
\begin{equation}
[\theta X, \theta Y] = (S^{-1} \bar{X} S) (S^{-1} \bar{Y} S)
- (S^{-1} \bar{Y} S)(S^{-1} \bar{X} S) =
S^{-1} \overline{(XY-YX)} S =\theta[X,Y].
\end{equation}
Thus we need the following
to complete the argument that $SU(N) = KAK$ of Definition \ref{def:KAK}
is well-defined: 
(i) $\mathfrak{a} \subset \mathfrak{p}$,
(ii) $\mathfrak{a}$ is commutative, and
(iii) no larger subalgebra of $\mathfrak{p}$ containing $\mathfrak{a}$
is commutative.

\begin{lemma}   
\label{lem:S_props}
Let $S=(-i\sigma^y_1)(-i\sigma^y_2)\cdots(-i\sigma^y_n)$
be as in Definition \ref{def:KAK}.  Then
(i) $S \ket{j} = (-1)^{\# j} \ket{N-j-1}$, (ii)
$\bra{j} S = (-1)^{n-\#j} \bra{N-j-1}$, and (iii)
$S \sigma^x_j = - \sigma^x_j S$, $S \sigma^y_j =
\sigma^y_j S$, and $S \sigma^z_j = -\sigma^z_j S$.
Note that (ii) refers to a composition of linear maps.
\end{lemma}

\begin{sketch}
For (i), compute.  For (ii), consider $\bra{j} S \ket{k}$ for
$\ket{k}$ varying over all computational basis states.  Then apply
(i) for (ii).  For (iii), nonlike Pauli matrices anticommute, while
$S$ itself is a tensor.
\end{sketch}

\begin{lemma}  Let $\mathfrak{a}$ be as in Definition \ref{def:KAK}.
Then $\mathfrak{a} \subset \mathfrak{p}$.
\end{lemma}

\begin{proof}  There are two coordinate computations to complete in this
case.  For the first, momentarily extend the definition of $\theta$ to
$\tilde{\theta}$ acting on $\mathfrak{u}(N)$ by the same formula.  Then
\begin{displaymath}
\begin{array}{lcl}
\tilde{\theta} [\ i \ket{j} \bra{j} + i \ket{N-j-1}\bra{N-j-1} \ ] & = & 
(-1)^n S (\ -i \ket{j} \bra{j} - i \ket{N-j-1} \bra{N-j-1} \ ) S \\
& = & (-1)^{n+1} i [\ (-1)^n \ket{N-j-1} \bra{N-j-1} + 
(-1)^n \ket{j} \bra{j} \ ] \\
& = & - i \ket{j} \bra{j} - i \ket{N-j-1} \bra{N-j-1} \\
\end{array}
\end{displaymath}
Thus $i \ket{j} \bra{j} + i \ket{N-j-1}\bra{N-j-1}$ is in the $-1$-eigenspace
of $\tilde{\theta}$, 
so that the elements of the first
set of the definition of $\mathfrak{a}$ are contained
in $\mathfrak{p}$.  For the second basis set in case $n$ even,
\begin{displaymath}
\begin{array}{lcl}
\theta  [\ i \ket{j} \bra{N-j-1} 
+ i \ket{N-j-1} \bra{j} \ ] & = & 
S [\ (-i) \ket{j} \bra{N-j-1} 
+ (-i)\ket{N-j-1} \bra{j} \ ] S \\
& = & [\ (-1)^{\#j + [n - (n - \#j)]} (-i) \ket{N-j-1} \bra{j}
\\
& & \quad \quad  \ +  \ (-1)^{n-\#j+(n-\#j)} (-i) \ket{j}\bra{N-j-1} \ ] \\
& = & (-i) \ket{N-j-1} \bra{j} +(-i) \ket{j} \bra{N-j-1} 
\end{array}
\end{displaymath} 
Thus $i \ket{j} \bra{N-j-1} + i \ket{N-j-1} \bra{j} \in \mathfrak{p}$,
in case $n$ even.
\end{proof}

\begin{proposition}
Recall $\mathfrak{a}$ from Definition \ref{def:KAK}.  Then
$\mathfrak{a}$ is commutative.
\end{proposition}

\begin{proof}
Throughout, $0 \leq j,k \leq N/2-1$, $N=2^n$.  The following three
computations of Lie brackets suffice.
\begin{displaymath}
\begin{array}{l}
\big[ 
\ \ i \ket{j} \bra{j} + i \ket{N-j-1} \bra{N-j-1}\ \ , \ \ i \ket{k} \bra{k} + 
i \ket{N-k-1} \bra{N-k-1} \ \ \big]  = \\
\quad 
- \ket{j} \langle j | k \rangle \bra{k} - \ket{N-j-1} \langle N-j-1 | N-k-1
\rangle \bra{N-k-1} + \\
\quad \quad \ket{k} \langle k | j \rangle \bra{j} +
\ket{N-k-1} \langle N-k-1 | N-j-1 \rangle \bra{N-j-1} \\
\\
\big[ \ \ (-i)^{n+1} \ket{j} \bra{N-j-1} + i^{n-1} \ket{N-j-1} \bra{j}\ \ 
, \ \ (-i)^{n+1} \ket{k} \bra{N-k-1} +
 i^{n-1} \ket{N-k-1} \bra{k} \ \ \big]  = \\
\quad - \ket{j} \langle N-j-1 | N-k-1 \rangle \bra{k} 
\ - \ket{N-j-1} \langle j | k \rangle \bra{k} \\
\quad \quad +\ket{k} \langle N-k-1 | N-j-1 \rangle \bra{j} \ \ +
\ket{N-k-1} \langle k | j \rangle \bra{N-j-1} \\
\\
\big[ \ \ i \ket{j} \bra{j} + i \ket{N-j-1} \bra{N-j-1} \ \ ,
\ \ (-i)^{n+1} 
\ket{k} \bra{N-k-1} + i^{n-1} \ket{N-k-1} \bra{k} \ \ \big] = \\
\quad 
(-i)^n \ket{j} \langle j | k \rangle \bra{N-k-1} 
\ + \ i^n \ket{N-j-1} \langle N-j-1 | N-k-1 \rangle \bra{k} \\
\quad \quad 
-(-i)^n \ket{k} \langle N-k-1 | N-j-1 \rangle \bra{N-j-1} \ -\ i^n 
\ket{N-k-1} \langle k | j \rangle \bra{j} \\
\end{array}
\end{displaymath}
Each of the final expressions is zero in case $j \neq k$ and also
zero in case $j=k$.  Thus, $\mathfrak{a}$ is commutative.
\end{proof}

The arguments above almost complete the proof that the {\tt CCD}
$SU(N)=KAK$ is well-defined.  In the abstract, one also needs a fairly large
coordinate computation which verifies $\mathfrak{a}$ is maximal
commutative.  This would verify that for any
$X \in \mathfrak{p}$ with $[X,H]=0$ for all $H \in \mathfrak{a}$,
one must in fact have $X \in \mathfrak{a}$.

Rather than complete that task, we instead appeal to the Cartan
classification \cite[pg.518,tableV]{Helgason:01}.  Ostensibly
a classification of globally symmetric spaces, this classification
also describes all possible Cartan involutions of any real semisimple
group $G$ up to Lie isomorphism.  For $G=SU(N)$, there are three
overall possibilities grouped as type {\bf AI}, {\bf AII},
and {\bf AIII}.  For each, the \emph{rank} refers to the dimension of
any maximal commutative subalgebra $\mathfrak{a}$ of $\mathfrak{p}$.
This dimension may not vary by subalgebra, since any two such $\mathfrak{a}_1$,
$\mathfrak{a}_2$ must have $k \mathfrak{a}_1 k^{-1}=\mathfrak{a}_2$
for some $k \in K$.  We now excerpt from the table the possibilities
for $G=SU(N)$:

\begin{center}
\begin{tabular}{|c|c|c|c|}
\hline
type & domain $\mathfrak{g}$
of $\theta: \mathfrak{g} \rightarrow \mathfrak{g}$
& isomorphism representative of $K$ & rank \\
\hline
\hline
{\bf AI} & $\mathfrak{su}(N)$ & $SO(N)$ & $N-1$ \\
\hline
{\bf AII} & $\mathfrak{su}(N)$ & $Sp(N/2)$ & $N/2-1$ \\
\hline
{\bf AIII} & $\mathfrak{su}(N)$ & $S[U(p)\oplus U(q)], p+q=N$ 
& $\mbox{min}(p,q)$ \\
\hline
\end{tabular}
\end{center}

Suppose then for the moment that the number of qubits $n$ is even.
No type {\bf AIII} Cartan involution admits an
$\mathfrak{a}$ of dimension $N-1$.  Indeed, if $p+q=N$, then
$\mbox{min} (p,q)\leq N/2 < N-1$.  The same is true of type
{\bf AII} involutions, i.e. $N-1>N/2-1$.  Hence we see that
$A$ must be maximal, and for $n$ even the Cartan involution
$\theta$ must have type {\bf AI}.

What remains is to prove that $\mathfrak{a}$ is maximal in $\mathfrak{p}$
in case $n$ odd.  This follows by a dimension count \emph{if}
the Cartan involution is type {\bf AII}.  We thus postpone noting
this point until after the proof of Theorem \ref{thm:K_sym}.
See Remark \ref{rem:AII_a}.

As an aside, type {\bf AIII} involutions do not appear in this work
but have been used in quantum circuit design.
Indeed, the {\tt CS}-decomposition \cite{Tucci:99,Golub:96}
is an example of a $KAK$ decomposition arising from a type
{\bf AIII} involution.  Elements within the appropriate $K$ group may be
interpreted as products of computations on the last $n-1$ lines with
computations on these lines controlled on the first qubit.

\subsection*{Entanglers}

In the two-qubit case, the following computation $E$ has the following
property:
\begin{equation}
E = (1/\sqrt{2})
\left(
\begin{array}{rrrr}
1 & i & 0 & 0 \\
0 & 0 & 1 & i \\
0 & 0 & -1 & i \\
1 & -i & 0 & 0 \\
\end{array}
\right) \mbox{ satisfies } E^\dagger [SU(2) \otimes SU(2)] E = SO(4)
\end{equation}
Using more Lie theory terminology, recall
the adjoint representation of $G$ on $\mathfrak{g}$ given by
$\mbox{Ad}(g)[X] = g X g^{-1}$.  Then we may restate
$\{\mbox{Ad}(E^\dagger)\} [SU(2) \otimes SU(2)] = SO(4)$.  This provides a
physical interpretation for the low dimensional
isomorphism $\mathfrak{su}(2) \oplus \mathfrak{su}(2)
\cong \mathfrak{so}(4)$.
We would like entanglers for the
concurrence canonical decomposition.

\begin{definition}
Let $\theta_{{\bf AI}}: \mathfrak{su}(2^n) \rightarrow \mathfrak{su}(2^n)$
denote the usual type {\bf AI} Cartan involution
$\theta_{{\bf AI}} (X) = \bar{X}$ associated to $SO(N) \subset SU(N)$.
We say $E \in SU(2^n)$ is an entangler iff
the following diagram commutes:
\begin{equation}
\begin{array}{rcccl}
& \mathfrak{su}(N) & {\buildrel {\theta_{{\bf AI}}} \over \longrightarrow}
& \mathfrak{su}(N) & \\
\mbox{Ad}(E) & \downarrow & & \downarrow & \mbox{Ad}(E) \\
& \mathfrak{su}(N) & 
{\buildrel {\theta} \over \longrightarrow} & 
\mathfrak{su}(N) \\
\end{array}
\end{equation}
In particular as both groups are connected,
we must have $\mbox{Ad}(E)[SO(N)]=K$.
\end{definition}

We next prove the surprising fact that there are no entanglers when
$n$ is odd.  For this, we need to recall the central subgroup
$\mathcal{Z}[SU(N)] 
= \{ v \in SU(N) \; ; \; v u v^\dagger = u \mbox{ for all } u \in SU(N)\}$.  
The center is in fact the set of all phase computations corresponding
to the $N^{\mbox{th}}$ roots of unity:
\begin{equation}
\mathcal{Z}[SU(N)] =
\{ \xi {\bf 1} \; ; \; \xi^N = 1 \} \quad
\mbox{\cite[pg.310,516]{Helgason:01}}
\end{equation}
With this fact recalled from the literature, we have the following
lemma.

\begin{lemma}
\label{lem:center}
Suppose that for $v \in SU(N)$, $[\mbox{Ad}(v)](X)=vXv^\dagger = X$ for every
$X \in \mathfrak{su}(N)$.  Then $v = \xi {\bf 1}$ for some
$\xi \in \mathbb{C}$ with $\xi^N = 1$.  (Hence
$\xi = \mbox{e}^{2 \pi i k / N}, 0 \leq k \leq N-1$.)
\end{lemma}

\begin{proof}
Recall that $\mbox{exp}: \mathfrak{su}(N) \rightarrow SU(N)$ is onto.
Thus each $u \in SU(N)$ may be written as $\mbox{exp} X$ for some
$X$.  Thus, consider the 
\hbox{one}-parameter-subgroup \cite[pg.104]{Helgason:01}
$\gamma:\mathbb{R} \rightarrow SU(N)$
given by
\hbox{$t \mapsto v [\mbox{exp}(tX)] v^\dagger$}.  This has derivative
$\frac{d\gamma}{dt}|_{t=0}=vXv^\dagger=X$, and by uniqueness of 
one-parameter-subgroups \cite[pg.103,Cor.1.5]{Helgason:01}
$v \mbox{exp}(tX) v^\dagger = \mbox{exp}(tX)$ for all $t$.  Taking $t=1$,
we see $v u v^\dagger = u$ for a generic $u \in SU(N)$.
\end{proof}

\begin{proposition}
If the number of qubits $n$ is odd, then 
there does not exist an entangler $E \in U(N)$.
\end{proposition}

\begin{proof}
Assume by way of contradiction that there does exist an entangler $E$
for $n$ odd.  Then for all $X \in \mathfrak{su}(N)$, we have the
following equation.
\begin{equation}
\label{eq:cantwork}
(E E^T) \bar{X} (E E^T)^\dagger \ = \
E \; \theta_{{\bf AI}}  [E^\dagger X E] \; E^\dagger \ = \ \theta (X) 
\; = \; S \bar{X} S^{-1}
\end{equation}
Since we may vary $Y=\bar{X}$ over $\mathfrak{su}(N)$ as well,
this implies that $S^{-1} E E^T$ satisfies the hypothesis of
Lemma \ref{lem:center}.
Thus $S^{-1} E E^T = \xi {\bf 1}$ for $\xi^N = 1$
or $E E^T = (\xi {\bf 1}) S$.  {Contradiction},
for $E E^T$ is always a complex symmetric matrix while $(\xi {\bf 1})S$ 
is not a complex symmetric matrix when $n$ is odd.
\end{proof}

\begin{scholium}
\label{scho:entform}
For an even number of qubits $n$, the matrix
$E \in U(N)$ is an entangler iff $E E^T = (\xi {\bf 1}) S$,
where $\xi^N=1$.
\end{scholium}

There are many possible entanglers.  Indeed, even in two-qubits
other choices have been used
\cite{Makhlin:00,Zhang:03}.  One possibility given
$n$ even is to take the $n/2$ fold tensor product
$E \otimes E \otimes \cdots \otimes E$.  However, we prefer the
following choice as a standard instead, since it highlights the
mapping of the computational basis to Greenberger-Horne-Zeilinger
states.

\begin{definition}
\label{def:stdE}
Suppose $n$ is even, and
write $S=(-i \sigma^y_1)(-i \sigma^y_2)\cdots
(-i \sigma^y_n) = \sum_{j=0}^{N/2-1}
\epsilon_j (\ket{j}\bra{N-j-1}+\ket{N-j-1}\bra{j})$, with 
$\epsilon_j = (-1)^{\# j}$,
where $\# j$ is the number of $1$'s in the binary expression for $j$.  The 
standard entangler $E_0$ in $n$-qubits is then given by
\begin{equation}
E_0 = \frac{1}{\sqrt{2}}
\sum_{j=0}^{N/2-1} \ket{j}\bra{2j}+i \ket{j}\bra{2j+1}
+\epsilon_j (\ket{N-j-1}\bra{2j}-i \ket{N-j-1}\bra{2j+1})
\end{equation}
\end{definition}

\begin{proposition}  The standard entangler $E_0$ is an
entangler.
\end{proposition}

\begin{proof}
First, we omit due to reasons of space a set of row operations which
verifies that $\mbox{det}(E_0)=1$.  Then
we may write out an expression for $E_0^T$ by reversing the indices in
each bra-ket pair:
\begin{equation}
E_0^T = \frac{1}{\sqrt{2}}
\sum_{k=0}^{N/2-1} \ket{2k}\bra{k}+i \ket{2k+1}\bra{k}
+\epsilon_k (\ket{2k}\bra{N-k-1}-i \ket{2k+1}\bra{N-k-1})
\end{equation}
Then Scholium \ref{scho:entform} shows that the following computation
suffices to prove that $E$ is an entangler.
\begin{equation}
\begin{array}{lcl}
E_0 E_0^T & = &
\frac{1}{2} \sum_{j=0}^{N/2-1} \ket{j}\bra{j}\; + \; \epsilon_j \ket{j}\bra{N-j-1} 
\; + \; i^2 \ket{j}\bra{j} +  \epsilon_j \ket{j}\bra{N-j-1} \\
& & \; + \; \epsilon_j \ket{N-j-1}\bra{j}+\epsilon_j^2 \ket{N-j-1}\bra{N-j-1} \\
& &  \; + \; 
\epsilon_j (\ket{N-j-1}\bra{j}+i^2\epsilon_j^2\ket{N-j-1}\bra{N-j-1})\\ 
& = & \sum_{j=0}^{N/2-1} \epsilon_j (\ket{j}\bra{N-j-1}+
\ket{N-j-1}\bra{j}) \\
& = & (-i \sigma^y_1)(-i \sigma^y_2) \cdots (-i \sigma^y_n) \\
\end{array}
\end{equation}
This concludes the coordinate computation.
\end{proof}

In the next section, we will also make use of the following lemma.
The computation is similar.

\begin{lemma}
\label{lem:isreal}
$E_0^T E_0$ is diagonal and real.  In fact,
$E_0^T E_0 = \ket{0}\bra{0}-\ket{1}\bra{1}+\ket{2}\bra{2}-\ket{3}\bra{3}+
\cdots$
\end{lemma}

\begin{proof}
Computing the reversed product:
\begin{equation}
\begin{array}{lcl}
E_0^T E_0 & = & \frac{1}{2}
\sum_{j=0}^{N/2-1} \ket{2j}\bra{2j}\; + \; i\ket{2j+1}\bra{2j}
\; + \; i\ket{2j}\bra{2j+1} \; - \; \ket{2j+1}\bra{2j+1}   \\
& & \; + \; \epsilon_j^2 \ket{2j}\bra{2j}-i\epsilon_j^2\ket{2j}\bra{2j+1}-i\epsilon_j^2
\ket{2j+1}\bra{2j}+i^2\epsilon_j^2 \ket{2j+1}\bra{2j+1} \\
& = & \sum_{j=0}^{N/2-1} \ket{2j}\bra{2j}-\ket{2j+1}\bra{2j+1} \\
\end{array}
\end{equation}
This concludes the proof.
\end{proof}

\begin{example}
Although this example is large, we explicitly describe the 
standard four-qubit entangler.
\begin{equation}
E_0 = 
(1/\sqrt{2})\left(
\begin{array}{cccccccccccccccc}
1 & i & 0 & 0 & 0 & 0 & 0 & 0 & 0 & 0 & 0 & 0 & 0 & 0 & 0 & 0 \\
0 & 0 & 1 & i & 0 & 0 & 0 & 0 & 0 & 0 & 0 & 0 & 0 & 0 & 0 & 0 \\
0 & 0 & 0 & 0 & 1 & i & 0 & 0 & 0 & 0 & 0 & 0 & 0 & 0 & 0 & 0 \\
0 & 0 & 0 & 0 & 0 & 0 & 1 & i & 0 & 0 & 0 & 0 & 0 & 0 & 0 & 0 \\
0 & 0 & 0 & 0 & 0 & 0 & 0 & 0 & 1 & i & 0 & 0 & 0 & 0 & 0 & 0 \\
0 & 0 & 0 & 0 & 0 & 0 & 0 & 0 & 0 & 0 & 1 & i & 0 & 0 & 0 & 0 \\
0 & 0 & 0 & 0 & 0 & 0 & 0 & 0 & 0 & 0 & 0 & 0 & 1 & i & 0 & 0 \\
0 & 0 & 0 & 0 & 0 & 0 & 0 & 0 & 0 & 0 & 0 & 0 & 0 & 0 & 1 & i \\
0 & 0 & 0 & 0 & 0 & 0 & 0 & 0 & 0 & 0 & 0 & 0 & 0 & 0 & -1 & i \\
0 & 0 & 0 & 0 & 0 & 0 & 0 & 0 & 0 & 0 & 0 & 0 & 1 & -i & 0 & 0 \\
0 & 0 & 0 & 0 & 0 & 0 & 0 & 0 & 0 & 0 & 1 & -i & 0 & 0 & 0 & 0 \\
0 & 0 & 0 & 0 & 0 & 0 & 0 & 0 & -1 & i & 0 & 0 & 0 & 0 & 0 & 0 \\
0 & 0 & 0 & 0 & 0 & 0 & 1 & -i & 0 & 0 & 0 & 0 & 0 & 0 & 0 & 0 \\
0 & 0 & 0 & 0 & -1 & i & 0 & 0 & 0 & 0 & 0 & 0 & 0 & 0 & 0 & 0 \\
0 & 0 & -1 & i & 0 & 0 & 0 & 0 & 0 & 0 & 0 & 0 & 0 & 0 & 0 & 0 \\
1 & -i & 0 & 0 & 0 & 0 & 0 & 0 & 0 & 0 & 0 & 0 & 0 & 0 & 0 & 0 \\
\end{array}
\right)
\end{equation}
Note that the antidiagonal pattern mirrors $S=
(-i\sigma^y_1)(-i \sigma^y_2) \cdots (-i \sigma^y_n)$
and that each computational basis state maps to a relative phase of
a {\tt GHZ} state.
\end{example}

\section*{Finaglers}

There do not exist entanglers when the number of qubits $n$ is
odd, because $\mathfrak{k} \cong \mathfrak{sp}(N/2)$ rather than
$\mathfrak{so}(N)$.  Cf. the as yet unproven Theorem \ref{thm:K_sym}.  
Yet the fairly abstract embedding $K$ of
$Sp(N/2)$ into $SU(N)$ might be made more standard.  This
is indeed possible, and we call the any matrix which rotates $K$ to the
standard $Sp(N/2)$ a finagler.

\vbox{
\begin{definition}
Let $\theta_{{\bf AII}}:\mathfrak{su}(N) \rightarrow \mathfrak{su}(N)$
be the standard Cartan involution \cite[pg.445]{Helgason:01}
fixing $\mathfrak{sp}(N/2)$,
i.e. $\theta_{{\bf AII}} (X) = (-i \sigma^y \otimes {\bf 1}_{N/2})
X^T (-i \sigma^y \otimes {\bf 1}_{N/2}) =
(-i \sigma^y \otimes {\bf 1}_{N/2})^{-1}
\bar{X} (-i \sigma^y \otimes {\bf 1}_{N/2})$.
Then a finagler $F$ is any $F\in SU(2^n)$ which causes the following
diagram to commute:
\begin{equation}
\begin{array}{rcccl}
& \mathfrak{su}(N) & {\buildrel {\theta_{{\bf AII}}} \over \longrightarrow}
& \mathfrak{su}(N) & \\
\mbox{Ad}(F) & \downarrow & & \downarrow & \mbox{Ad}(F) \\
& \mathfrak{su}(N) & 
{\buildrel {\theta} \over \longrightarrow} & 
\mathfrak{su}(N) \\
\end{array}
\end{equation}
If $F \in SU(N)$, then we say $F$ finagles iff $F$ is a finagler.
\end{definition}
}

\begin{proposition}
$F$ is a finagler iff 
$F (-i \sigma^y \otimes {\bf 1}_{N/2})^T F^T = 
(\xi {\bf 1})
(-i \sigma^y_1)(-i \sigma^y_2) \cdots (-i\sigma^y_n) = (\xi{\bf 1})S$,
$\xi^N = 1$.
\end{proposition}

\begin{proof}
For convenience, label $\Sigma = -i \sigma^y \otimes {\bf 1}_{N/2}$.
$(F \mbox{ finagles})$ $\Longleftrightarrow$
$[F \Sigma^{-1} (\overline{F^\dagger X F}) \Sigma F^\dagger = S^{-1} \bar{X} S
\ \forall X \in \mathfrak{su}(N)]$
$\Longleftrightarrow$ $[F \Sigma^T F^T = (\xi {\bf 1})S, \xi^N=1]$.
Note that the second equivalence uses Lemma \ref{lem:center}.
\end{proof}

\begin{example}  In three qubits, we see the following computation is
a finagler by direct computation.
\begin{equation}
F = (1/\sqrt{2})
\left(
\begin{array}{rrrrrrrr}
1 & 0 & 0 & 0 & 1 & 0 & 0 & 0 \\
0 & 1 & 0 & 0 & 0 &-1 & 0 & 0 \\
0 & 0 & 1 & 0 & 0 & 0 &-1 & 0 \\
0 & 0 & 0 & 1 & 0 & 0 & 0 & 1 \\
0 & 0 & 0 & 1 & 0 & 0 & 0 &-1 \\
0 & 0 & 1 & 0 & 0 & 0 & 1 & 0 \\
0 & 1 & 0 & 0 & 0 & 1 & 0 & 0 \\
1 & 0 & 0 & 0 &-1 & 0 & 0 & 0 \\
\end{array}
\right)
\end{equation}
Unlike entanglers, it is possible for $F=\bar{F}$.
The finagler maps computational basis states to 
{\tt GHZ} states.
\end{example}

\begin{definition}
\label{def:stdF}
Fix $n$ an odd number of qubits.  Let
$S = (-i\sigma^y_1)(-i\sigma^y_2)\cdots(-i\sigma^y_n) = \sum_{j=0}^{N/2-1}
\iota_j (\ket{N-j-1}\bra{j}-\ket{j}\bra{N-j-1})$ with
$\iota_j = \pm 1$.  The standard finagler $F_0$ is defined to be
the following linear operator:
\begin{equation}
F_0 = \sum_{j=0}^{N/2-1} 
\ket{j}\bra{j}  +  \ket{N-j-1}\bra{j}
\; + \; \iota_j ( \ket{j}\bra{N/2+j} - \ket{N-j-1}\bra{N/2+j})
\end{equation}
Note that the standard finagler is real.
\end{definition}

\begin{proposition}
The standard finagler $F_0$ finagles.
\end{proposition}

\vbox{
\begin{proof}
We again omit the column operations verifying $\mbox{det}(F_0)=1$,
as this would take several pages.
Thus, let  $\Sigma = -i \sigma^y \otimes {\bf 1}_{N/2}$ be expanded as
$\Sigma = \sum_{j=0}^{N/2-1} \ket{j}\bra{N/2+j}-\ket{N/2+j}\bra{j}$.
We have the following equation:
\begin{equation}
F_0 \Sigma  =  \frac{1}{\sqrt{2}}\sum_{j=0}^{N/2-1} 
\ket{j}\bra{N/2+j} \; + \; \ket{N-j-1}\bra{N/2+j} \; - \;
\iota_j (\ket{j}\bra{j} - \ket{N-j-1}\bra{j})
\end{equation}
Moreover, $F_0^T = \frac{1}{\sqrt{2}}\sum_{j=0}^{N/2-1} 
\ket{j}\bra{j}  +  \ket{j}\bra{N-j-1}
\; + \; \iota_j ( \ket{N/2+j}\bra{j} - \ket{N/2+j}\bra{N-j-1})$.
Thus we see that
\begin{equation}
\begin{array}{lcl}
(F_0 \Sigma) F_0^T & = & \frac{1}{2}\sum_{j=0}^{N/2-1}
\iota_j (\ket{j}\bra{j}-\ket{j}\bra{N-j-1}+
\ket{N-j-1}\bra{j}-\ket{N-j-1}\bra{N-j-1}) \\
& & \quad \quad - \; 
\iota_j (\ket{j} \bra{j} +\ket{j}\bra{N-j-1}-\ket{N-j-1}\bra{j}
-\ket{N-j-1}\bra{N-j-1}) \\
& = & \sum_{j=0}^{N/2-1} \iota_j (\ket{N-j-1}\bra{j}-\ket{j}\bra{N-j-1})
\\
\end{array}
\end{equation}
This concludes the proof.
\end{proof}
}

We also briefly review how $Sp(N/2)$ embeds into $SU(N)$.  By one
standard definition of the group \cite[pg.446]{Helgason:01},
\begin{equation}
\label{eq:Hel_sp}
\mathfrak{sp}(N/2) = \Bigg\{
\left(
\begin{array}{rr}
X_1 & X_2 \\
X_3 & -X_1^T \\
\end{array}
\right) \ ; \ X_j = \bar{X}_j, X_{2,3} \mbox{ symmetric} \Bigg\}
\end{equation}
Another standard definition \cite[pp.34-36]{Knapp:96} uses a symmetry
in matrices of quaternions.  Note that the matrices of
Equation \ref{eq:Hel_sp} are not elements of $\mathfrak{su}(N)$.
Rather, the
$+1$ eigenspace of $\theta_{{\bf AII}}(X) = \Sigma^{-1} \bar{X} \Sigma$ is:
\begin{equation}
\mathfrak{sp}(N/2)
= \Bigg\{
\left(
\begin{array}{rr}
V & W \\
-W^\dagger & \bar{V}  \\
\end{array}
\right)
\ ; \ 
V \in \mathfrak{u}(N/2), W=W^T \mbox{ is complex symmetric} \Bigg\}
\end{equation}
(For example, $Sp(4) \subset SU(8)$ this is $36$ dimensional.  For $W$ includes
two real symmetric matrices with $10$ dimensions each, while
$\mathfrak{u}(4)$ is $16$ dimensional.)
One may verify this is also a copy of $\mathfrak{sp}(N/2)$, so that
$\mathfrak{k}$ is a copy of $\mathfrak{sp}(N/2)$ as well.
Also, note that for $k \in K$, in particular $k \in \otimes_1^n SU(2)$,
we expect $F k F^\dagger$ to be in the \emph{copy of $Sp(N/2)$ above}
rather than to be a real matrix in an orthogonal subgroup of
$SU(N)$.  Finally, note that the exponentiating the Lie algebra above
is not the best way to write out a closed form for elements of the
global group $Sp(N/2)$.  Rather, we have a block form:
\begin{equation}
Sp(N/2) = \{ V \in SU(N) \; ; \; V^T \Sigma V = \Sigma \} =
\Bigg\{
\left(
\begin{array}{rr}
A & B \\
C & D \\
\end{array}
\right) \in SU(N)
\; ; \; 
\begin{array}{l}
A^T C \mbox{ is symmetric}, B^T D \mbox{ is symmetric}, \\
A^TD-C^TB={\bf 1}
\end{array}
\Bigg\}
\end{equation}

\subsection*{$K$ is the symmetry group of the concurrence form}

We are now in a position to provide the physical interpretation of
$K$.  Namely, $K$ is the symmetry group of the concurrence bilinear
form, as stated in Theorem \ref{thm:K_sym}.

\begin{proof}[\ of {\bf Theorem} \ref{thm:K_sym}]
We first prove that $v \in K$ iff $\mathcal{C}_n(v\ket{\phi},v\ket{\psi})=
\mathcal{C}_n(\ket{\phi},\ket{\psi})$ for all $\ket{\phi}$,
$\ket{\psi} \in \mathcal{H}_n$.
Let $X = \log \; v$.  
Since $X \in \mathfrak{su}(N)$, $X$ is anti-Hermitian,
i.e. $X=-X^\dagger=-\bar{X}^T$.  Finally, recall $S = (-i \sigma^y_1)(-i \sigma^y_2)
\cdots (-i \sigma^y_n)$.  Thus in mathematical notation, we have
for $w,x \in \mathcal{H}_n$ the concurrence form given by
$\mathcal{C}_n (w,x)= w^T S v$.
Hence
\begin{equation}
(X=S^{-1} \bar{X} S) \Longleftrightarrow
(SX=\bar{X}S) \Longleftrightarrow
(SX=-X^T S) \Longleftrightarrow
(X^T S + SX = 0) \Longleftrightarrow
({v}^T S v = S)
\end{equation}
Now the first item is equivalent to $v \in K$
while the last is equivalent to $\mathcal{C}(vw,vx)=
(w^T v^T) S (v x) = w^T (v^T S v) x = w^T S x = \mathcal{C}(w,x)$
for all $w,x \in \mathcal{H}_n$.

We next prove that for $n$ odd, $K \cong Sp(N/2)$.
To do so, it suffices to show $n$ odd implies $\mathcal{C}_n$ is
a nondegenerate two-form on $\mathcal{H}_n$.
We first show $\mathcal{C}_n(x,w)=
-\mathcal{C}_n(w,x)$ for any $w,x \in \mathcal{H}_n$.
Noting that the transpose of a $1 \times 1$ matrix is again the same
matrix, we realize that $\mathcal{C}_n$ is a two-form as follows:
\begin{equation}
\mathcal{C}_n(w,x) = w^T S x = [w^T S x]^T = x^T S^T w =
-x^T S w = - \mathcal{C}_n(x,w)
\end{equation}
Moreover, consider the tensor expression for $S$.  We see that no
eigenvalues of $S$ are zero, and hence the form is nondegenerate. 
Thus, we must have $K \cong Sp(N/2)$.

Suppose now that $n$ is even.  We finally prove $K \cong SO(N)$.
It suffices to construct a real vector
space $V_{\mathbb{R}} \subset \mathcal{H}_n$ so that the following
properties hold:
\begin{itemize}
\item $K \cdot V_{\mathbb{R}} \subseteq V_{\mathbb{R}}$
\item The restriction of
$\mathcal{C}_n$ to $V_{\mathbb{R}} \times V_{\mathbb{R}}$ is 
the usual dot product in the coordinates of a given basis.
\end{itemize}
Consider then $V_{\mathbb{R}} = \mbox{span}_{ \bf \mathbb{R}}
\{ E_0 \ket{j} \; ; \; 0 \leq j \leq N-1 \}$, for $E_0$ the standard
entangler of Definition \ref{def:stdE}.  Since $E_0$ is an entangler, certainly
$K \cdot V_{\mathbb{R}} \subset V_{\mathbb{R}}$ since $K$ acts on this
real vector space by (real) orthogonal maps.  Moreover, consider the
concurrence on $V_{\mathbb{R}}$.  For $w,x$ in the $\mathbb{R}$ span
of the computational basis, we have $E_0w$, $E_0x$ generic vectors in
$V_{\mathbb{R}}$.  Then
\begin{equation}
\mathcal{C}_n(E_0 w, E_0 x) = (E_0 w)^T S (E_0 x) = w^T E_0^T S E_0 x =
w^T E_0^T E_0 E_0^T E_0 x = w^T {\bf 1} x = w\cdot x
\end{equation}
with the fourth equality by Lemma \ref{lem:isreal}. 
Hence in an even number of qubits, $K$ fixes a real inner product
on a real vector subspace of $\mathcal{H}_n$.  Thus
$K \cong SO(N)$.
\end{proof}

\begin{scholium}
\label{sch:C_n_and_dot}
For $n$ even, for $E_0$ the standard entangler of
Definition \ref{def:stdE},
for any $\ket{\phi}$, $\ket{\psi} \in \mathcal{H}_n$, 
we have $\mathcal{C}_n(E_0 \ket{\phi}, E_0 \ket{\psi}) =
\overline{\bra{\phi}} \psi \rangle$.
\end{scholium}

\begin{remark}
\label{rem:AII_a}
Note that independent of any discussion of the algebra $\mathfrak{a}$
in $n$ an odd number of qubits, we have shown that the Cartan involution
$\theta$ has type {\bf AII}.  Hence any commutative
$N/2-1$ dimensional subalgebra of $\mathfrak{p}$ must be maximal,
and the concurrence canonical decomposition $SU(N)=KAK$ is well-defined 
for $n$ odd.
\end{remark}

\begin{remark}
Similar to Scholium \ref{sch:C_n_and_dot}, note that the standard
(real) finagler $F_0$ of Definition \ref{def:stdF}
translates between the concurrence
and the more standard two-form $(w,x) \mapsto
w^T [(-i\sigma^y) \otimes {\bf 1}_{N/2}] x$.
Indeed, $F_0^{-1}=F_0^T$
since $F_0$ is orthogonal.  Moreover, let $w$, $x$ be in the real
span of the computational basis states $\{\ket{j} \; ; \; 0 \leq j \leq
N-1\}$.  Then we may view $\{ F_0 \ket{j} \; ; \; 0 \leq j \leq N-1\}$
as a finagled basis, and the pullback of the concurrence from
the finagled to the computational basis is the model two-form.
Indeed, labelling $\Sigma=(-i\sigma^y) \otimes {\bf 1}_{N/2}$,
$F_0 \Sigma F_0^T =S$ and $F_0$ real imply $\Sigma=F_0^T S F_0$.  Hence
\(\mathcal{C}_n(F_0 w,F_0 x) = (F_0 w)^T S (F_0 x) =
w^T  \Sigma x \).
\end{remark}

\subsection*{Cartan Involution in Coordinates}

We finally present the Cartan involution in coordinates and provide
some sample calculations.
Let $X \in \mathfrak{su}(N)$, say with
$X=\sum_{j,k=0}^{N-1} x_{j,k} \ket{k} \bra{j}$.
We now compute explicitly $\theta (X)$ so as to arrive
at coefficient expressions for $\mathfrak{p}$,
$\mathfrak{k}$.

\begin{displaymath}
\begin{array}{lclcl}
\theta (X) & = & (-1)^n \sum_{j,k=0}^{N-1}
\bar{x}_{j,k} S \ket{k} \bra{j} S & = (-1)^{n+n-\#k+\#j} \sum_{j,k=0}^{N-1}
\bar{x}_{j,k} \ket{N-k-1} \bra{N-j-1} \\
\end{array}
\end{displaymath}
Consequently, we have the following characterizations:
\begin{itemize}
\item $X=\sum_{j,k=0}^{N-1} x_{j,k} \ket{k} \bra{j} \in \mathfrak{p}$
iff $[(X \in \mathfrak{su}(N)) \mbox{ and } 
(x_{N-1-k,N-1-j}=(-1)^{\#j+\#k+1}\bar{x}_{k,j})]$

\item $X=\sum_{j,k=0}^{N-1} x_{j,k} \ket{k} \bra{j} \in \mathfrak{k}$
iff $[(X \in \mathfrak{su}(N)) \mbox{ and } 
(x_{N-1-k,N-1-j}=(-1)^{\#j+\#k}\bar{x}_{k,j})]$
\end{itemize}
This moreover produces the following description of $\mathfrak{k}$.
\begin{equation}
\label{eq:k_w_jokes}
\begin{array}{lcll}
\mathfrak{k}& = &
\mbox{span}_{\mathbb{R}}
& \{ \ket{k}\bra{j}-\ket{j}\bra{k}+(-1)^{\#j+\#k}\ket{N-k-1}\bra{N-j-1}
-(-1)^{\#j+\#k}\ket{N-j-1}\bra{N-k-1} \} \\
& & & \ \sqcup \ 
\{i\ket{k}\bra{j}+i\ket{j}\bra{k}+
(-1)^{\#k+\#j+1}i \ket{N-j-1}\bra{N-k-1}+(-1)^{\#j+\#k+1}i\ket{N-k-1}
\bra{N-j-1} \} \\
& & & \ \sqcup \ 
\{i\ket{j}\bra{j}-i\ket{N-j-1}\bra{N-j-1} \} \\
\end{array}
\end{equation}

\begin{remark}
We {\em warn} the reader that
the above expression does not allow one to count
dimensions.  Several repetitions occur from set to set, and moreover
the expressions may vanish in case $j=N-k-1$.
\end{remark}

\subsection*{Example in the two-qubit case}

Recall the subalgebra $[{\bf 1} \otimes \mathfrak{su}(2)] \oplus
[\mathfrak{su}(2)\otimes {\bf 1}]$ of infinitesimal transformations by
$SU(2) \otimes SU(2) \subseteq SU(4)$.  We show how the
above Equation \ref{eq:k_w_jokes} recovers this subalgebra in the
case of $n=2$ qubits.

We begin by plugging $k=0$, $j=1$.  Expanding into
binary (or writing out the matrix) makes clear this is a tensor,
and moreover a tensor by an identity matrix.  Recall again that
both are required to be in the Lie algebra of $SU(2)\otimes SU(2)$.
\begin{equation}
\begin{array}{lc}
\ket{0}\bra{1}-\ket{1}\bra{0}-\ket{3}\bra{2}+\ket{2}\bra{3} & = \\
\ket{00}\bra{01}-\ket{01}\bra{00}-\ket{11}\bra{10}+\ket{10}\bra{11} & = \\
(\ket{0}\bra{0}+\ket{1}\bra{1})\otimes (\ket{0}\bra{1}-\ket{1}\bra{0}) \\
\end{array}
\end{equation}
One may similarly analyze the following matrices:
\begin{equation}
\begin{array}{l}
i\ket{0}\bra{1}+i\ket{1}\bra{0}+i\ket{3}\bra{2}+i\ket{2}\bra{3} \\
\ket{0}\bra{2}-\ket{2}\bra{0}-\ket{3}\bra{1}+\ket{1}\bra{3} \\
i\ket{0}\bra{2}+i\ket{2}\bra{0}+i\ket{3}\bra{1}+i\ket{1}\bra{3} \\
\end{array}
\end{equation}
Note that for the next four expressions, substitution returns a $0$ matrix:
\begin{equation}
\begin{array}{l}
\ket{0}\bra{3}-\ket{3}\bra{0}+\ket{3}\bra{0}-\ket{0}\bra{3} \\
i\ket{0}\bra{3}+i\ket{3}\bra{0}-i\ket{3}\bra{0}-i\ket{0}\bra{3} \\
\ket{1}\bra{2}-\ket{2}\bra{1}+\ket{2}\bra{1}-\ket{1}\bra{2} \\
i\ket{1}\bra{2}+i\ket{2}\bra{1}-i\ket{2}\bra{1}-i\ket{1}\bra{2} \\
\end{array}
\end{equation}
Further substitution yields the following:
\begin{equation}
\begin{array}{l}
\ket{1}\bra{3}-\ket{3}\bra{1}+\ket{2}\bra{0}-\ket{0}\bra{2} \\
i\ket{1}\bra{3}+i\ket{3}\bra{1}+i\ket{2}\bra{0}+i\ket{0}\bra{2} \\
\ket{2}\bra{3}-\ket{3}\bra{2}-\ket{1}\bra{0}+\ket{0}\bra{1} \\
i\ket{2}\bra{3}+i\ket{3}\bra{2}+i\ket{1}\bra{0}+i\ket{0}\bra{1} \\
\end{array}
\end{equation}
Finally, we consider the diagonal matrices in $\mathfrak{k}$:
\begin{equation}
\begin{array}{l}
i\ket{0}\bra{0}-i\ket{3}\bra{3} \\
i\ket{1}\bra{1}-i\ket{2}\bra{2} \\
\end{array}
\end{equation}
Note that the $\mathbb{R}$ span of these two matrices coincides
with $\mathbb{R} (i\sigma_z^1) \oplus \mathbb{R} (i\sigma^2_z)$.

The Cartan involution formalism thus works, although in a cumbersome
way.  We next explore the answer it returns in the three-qubit
case.

\subsection*{Example in the three-qubit case, $K=Sp(4)$}

We now describe explicitly the output of Equation \ref{eq:k_w_jokes}
in three qubits.  The corresponding real Lie algebra is thirty-six
dimensional, which implies by the Cartan classification that
$K$ is an abstract copy of $Sp(4)$.  A copy of $SO(8)$ would rather
be twenty-eight dimensional.

The simplest way to organize the three qubit computation is to
appeal to separation.  We say a term $\ket{k}\bra{j}$ has
separation $|k-j|$ and extend linearly.  In Equation
\ref{eq:k_w_jokes}, each matrix described has a well-defined separation.
\begin{equation}
\begin{array}{ll}
\mbox{Separation 0} &
i \ket{0}\bra{0}-i\ket{7}\bra{7} \\
\mbox{Total 4} & i\ket{1}\bra{1}-i\ket{6}\bra{6} \\
& i\ket{2}\bra{2}-i\ket{5}\bra{5} \\
& i\ket{3}\bra{3}-i\ket{4}\bra{4} \\
\end{array}
\end{equation}

\begin{equation}
\begin{array}{ll}
\mbox{Separation 1} &
\ket{0}\bra{1}-\ket{1}\bra{0}-\ket{7}\bra{6}+\ket{6}\bra{7} \\
\mbox{Total 8} &
i\ket{0}\bra{1}+i\ket{1}\bra{0}+i\ket{7}\bra{6}+i\ket{6}\bra{7} \\
& \ket{1}\bra{2}-\ket{2}\bra{1}+\ket{6}\bra{5}-\ket{5}\bra{6} \\
& i\ket{1}\bra{2}+i\ket{2}\bra{1}-i\ket{6}\bra{5}-i\ket{5}\bra{6} \\
& \ket{2}\bra{3}-\ket{3}\bra{2}-\ket{5}\bra{4}+\ket{4}\bra{5} \\
& i\ket{2}\bra{3}+i\ket{3}\bra{2}+i\ket{5}\bra{4}+i\ket{4}\bra{5} \\
& \ket{3}\bra{4}-\ket{4}\bra{3} \\
& i\ket{3}\bra{4}+i\ket{4}\bra{3} \\
\end{array}
\end{equation}

\begin{equation}
\begin{array}{ll}
\mbox{Separation 2} & 
\ket{0}\bra{2}-\ket{2}\bra{0}-\ket{7}\bra{5}+\ket{5}\bra{7} \\
\mbox{Total 6} &
i\ket{0}\bra{2}+i\ket{2}\bra{0}+i\ket{7}\bra{5}+i\ket{5}\bra{7} \\
& \ket{1}\bra{3}-\ket{3}\bra{1}-\ket{6}\bra{4}+\ket{4}\bra{6} \\
& i\ket{1}\bra{3}+i\ket{3}\bra{1}+i\ket{6}\bra{4}+i\ket{4}\bra{6} \\
& \ket{2}\bra{4}-\ket{4}\bra{2}+\ket{5}\bra{3}-\ket{3}\bra{5} \\
& i\ket{2}\bra{4}+i\ket{4}\bra{2}-i\ket{5}\bra{3}-i\ket{3}\bra{5} \\
\end{array}
\end{equation}

\begin{equation}
\begin{array}{ll}
\mbox{Separation 3} &
\ket{0}\bra{3}-\ket{3}\bra{0}+\ket{7}\bra{4}-\ket{4}\bra{7} \\
\mbox{Total 6} &
i\ket{0}\bra{3}+i\ket{3}\bra{0}-i\ket{7}\bra{4}-i\ket{4}\bra{7} \\
& \ket{1}\bra{4}-\ket{4}\bra{1}+\ket{6}\bra{3}-\ket{3}\bra{6} \\
& i\ket{1}\bra{4}+i\ket{4}\bra{1}-i\ket{6}\bra{3}-i\ket{3}\bra{6} \\
& \ket{2}\bra{5}-\ket{5}\bra{2} \\
& i\ket{2}\bra{5}+i\ket{5}\bra{2} \\
\end{array}
\end{equation}

\begin{equation}
\begin{array}{ll}
\mbox{Separation 4} &
\ket{0}\bra{4}-\ket{4}\bra{0}-\ket{7}\bra{3}+\ket{3}\bra{7} \\
\mbox{Total 4} &
i\ket{0}\bra{4}+i\ket{4}\bra{0}+i\ket{7}\bra{3}+i\ket{3}\bra{7} \\
& \ket{1}\bra{5}-\ket{5}\bra{1}-\ket{6}\bra{2}+\ket{2}\bra{6} \\
& i\ket{1}\bra{5}+i\ket{5}\bra{1}+i\ket{6}\bra{2}+i\ket{2}\bra{6} \\
\end{array}
\end{equation}

\begin{equation}
\begin{array}{ll}
\mbox{Separation 5} &
\ket{0}\bra{5}-\ket{5}\bra{0}+\ket{7}\bra{2}-\ket{2}\bra{7} \\
\mbox{Total 4} &
i\ket{0}\bra{5}+i\ket{5}\bra{0}-i\ket{7}\bra{2}-i\ket{2}\bra{7} \\
&
\ket{1}\bra{6}-\ket{6}\bra{1} \\
& i\ket{1}\bra{6}+i\ket{6}\bra{1} \\
\end{array}
\end{equation}

\begin{equation}
\begin{array}{ll}
\mbox{Separation 6} &
\ket{0}\bra{6}-\ket{6}\bra{0}+\ket{7}\bra{1}-\ket{1}\bra{7} \\
\mbox{Total 2} &
i\ket{0}\bra{6}+i\ket{6}\bra{0}-i\ket{7}\bra{1}-i\ket{1}\bra{7} \\
\end{array}
\end{equation}

\begin{equation}
\begin{array}{ll}
\mbox{Separation 7} &
\ket{0}\bra{7}-\ket{7}\bra{0} \\
\mbox{Total 2} &
i\ket{0}\bra{7}+i\ket{7}\bra{0} \\
\end{array}
\end{equation}
Thus we see a total of $4+8+6+6+4+4+2+2=36$ real dimensions in
$\mathfrak{k}$.  Now by the Cartan classification
\cite[pg.518]{Helgason:01}, the Cartan involution $\theta$
must be either type {\bf AI} fixing an abstract copy of
$SO(28)$, type {\bf AIII} fixing some $S[U(p)\oplus U(q)]$ for
$p+q=2^n$, or else {\bf type AII} fixing an abstract copy of
$Sp(4)$.  Since only 
$Sp(4)$ is thirty-six dimensional, we see
$\mathfrak{k} \cong \mathfrak{sp}(4)$ and
$K\cong Sp(4)$.

\section{Applications to concurrence capacity}
\label{sec:capacity}

This section focuses on an application of the concurrence canonical
decomposition
$SU(N)=KAK$ of Definition \ref{def:KAK}
when the number of qubits $n$ is even.
Namely, we study how a given computation $v \in SU(N)$ may
change the concurrence of the quantum data state.  Since
we have the concurrence $C_n (\ket{\psi}) =
| \overline{\bra{\psi}} (-i\sigma^y_1)(-i\sigma^y_2)\cdots(-i\sigma^y_n)
\ket{\psi} |$ with the $n$-tangle $\tau_n = C_n^2$ for $n$
even, there are
immediate applications to the $n$-tangle as well.

Let $v \in SU(N)$.
Recall from Definition \ref{def:kappa}
that the concurrence capacity is defined as
\begin{equation}
\kappa(v)=\mbox{max} \{
C_n(v\ket{\psi}) \; ; \; {C_n(\ket{\psi})=0, \bra{\psi} \psi \rangle=1} \}
\end{equation}
Since we vary over all $C_n(\ket{\psi})=0$, we see that for $k \in K$
we have $\kappa(vk)=\kappa(v)$ by symmetry.  Immediately
$\kappa(kv)=\kappa(v)$.  Thus, for $v=k_1ak_2$ the {\tt C.C.} decomposition
of any $v \in SU(N)$, we have $\kappa(v)=\kappa(k_1ak_2)=\kappa(a)$.

We next describe the concurrence capacity of any $a \in A$.
The formalism makes strong use of entanglers to translate between
$\mathcal{C}_n$ and $(w,x) \mapsto w^Tx$.

\begin{definition}
\label{def:con_spec}
The \emph{concurrence spectrum} $\lambda_c(v)$ 
of $v \in SU(N)$ is the spectrum
of $E_0^\dagger v E_0 (E_0^\dagger v E_0)^T$, for $E_0$ the standard entangler of
Definition \ref{def:stdE}.  Note that the spectrum is the set of
eigenvalues since $\mathcal{H}_n$ is finite dimensional.
The \emph{convex hull} ${\tt CH}[\lambda_c(v)]$
of $\lambda_c(v)$ is the set of all line
segments joining all points of $\lambda_c(v)$, i.e.
\begin{equation}
{\tt CH}[\lambda_c(v)] = \Bigg\{ \sum_{z_j \in \lambda_c(v)} 
t_j z_j \ ; \ 0 \leq t_j \leq 1, \sum_{j=0}^{\#\lambda_c(v)} t_j =1 \Bigg\}
\end{equation}
\end{definition}

These definitions allow us then to prove the following general
results regarding concurrence capacity.  The techniques closely
follow those in prior work \cite{Zhang:03}.

\begin{lemma}
\label{lem:convex_hull}
Let $v \in SU(N)$, with {\tt CCD} $v=k_1 a k_2$ for 
$a=E_0 d E_0^\dagger$ for $d$ diagonal in $SU(N)$.
\begin{itemize}
\item $\lambda_c(v)=\lambda_c(a)=
\bigg\{ d_j^2 \; ; \; d = \sum_{j=0}^{N-1} d_j \ket{j}\bra{j} \bigg\}$.
\item $\kappa(v)=\kappa(a) = \mbox{max} \bigg\{
| \sum_{j=0}^{N-1} a_j^2 d_j^2 | 
\; ; \; \ket{\psi}= \sum_{j=0}^{N-1} a_j \ket{j},
\bra{\psi} \psi \rangle = 1, \overline{\bra{\psi}} \psi \rangle=0 \bigg\}$.
\item $(\kappa(v)=\kappa(a)=1)$ $\Longleftrightarrow$ 
$\big( 0 \in {\tt CH}[\lambda_c(v)]={\tt CH}[\lambda_c(a)] \big)$.
\end{itemize}
\end{lemma}

\begin{proof}
For the first item, recall $v=k_1 a k_2$.  Thus the following expression
results from expanding Definition \ref{def:con_spec}.
\begin{equation}
E_0^\dagger v E_0 (E_0^\dagger v E_0)^T=
[(E_0^\dagger k_1 E_0) (E_0^\dagger a E_0) (E_0^\dagger k_2 E_0)]
[(E_0^\dagger k_2 E_0)^T (E_0^\dagger a E_0)^T (E_0^\dagger k_1 E_0)^T]
\end{equation}
Label the elements of $SO(N)$ by $o_1 = (E_0^\dagger k_1 E_0)$,
$o_2=(E_0^\dagger k_2 E_0)$, and put $d=E_0^\dagger a E_0$ diagonal.  Then the 
above reduces to $o_1 d o_2 o_2^T d^T o_1^T = o_1 d^2 o_1^{-1}$, with spectrum
identical to $d$.

For the next item, compare the two-qubit case
\cite[Eq.(41)]{Zhang:03} and recall Scholium \ref{sch:C_n_and_dot}.
Suppose $C_n(\ket{\varphi})=0$.  Then per Scholium \ref{sch:C_n_and_dot},
for $\ket{\psi}=E_0^\dagger \ket{\varphi}$ we have
we see $0 = \mathcal{C}_n(E_0 E_0^\dagger \ket{\varphi}, E_0 E_0^\dagger \ket{\varphi})=
\overline{\bra{\psi}} \psi \rangle$.
Now for $\kappa(a)$, take $\ket{\psi}=E_0^\dagger \ket{\varphi}$.
We then maximize over expressions
$\mathcal{C}_n(a \ket{\varphi}, a \ket{\varphi})=
\mathcal{C}_n(E_0 E_{0}^\dagger a E_0 
\ket{\psi}, E_0 E_{0}^\dagger a E_0 \ket{\psi})=
\mathcal{C}_n(E_0 d \ket{\psi}, E_0 d \ket{\psi})=
\overline{\bra{\psi}} d^2 \ket{\psi}$.

The final item makes use of the Schwarz inequality.  For should 
the concurrence capacity be maximal, there is by 
compactness of the set of normalized
kets some normalized $\ket{\psi}$ with $C_n(\ket{\phi})=1$.
For $\ket{\psi}= \sum_{j=0}^{N-1} a_j \ket{j}$,
\begin{equation}
1=| \sum_{j=0}^{N-1} a_j^2 d_j^2 | \leq
\sum_{j=0}^{N-1} |a_j^2 d_j^2| = \sum_{j=0}^{N-1} |a_j|^2 = 1
\end{equation}
The Schwarz \emph{equality} further requires some
$z \in \mathbb{C}$, $z \bar{z}=1$, so that
$a_j^2 d_j^2 = |a_j|^2 z, \forall j$.  Now since 
$\overline{\bra{\psi}} \psi \rangle=0$,
\begin{equation}
0=\sum_{j=0}^{N-1} a_j^2 = \sum_{j=0}^{N-1} |a_j|^2 \bar{d}_j^2 z
\end{equation}
Multiplying through by $\bar{z}$ and taking the complex conjugate,
we see $0 \in {\tt CH}[\lambda_c(v)]$.
\end{proof}

As already noted in the introduction, the concurrence capacity
$\kappa$ is properly thought of as a function of $A$ rather than
a function of $SU(N)$.  This is advantageous from a computational standpoint, because in order to calculate $\kappa(v)$ one need minimize over a function involving $N-1$ real parameters in A versus $N^2-1$ parameters describing a general $v\in SU(N)$.   We next consider typical values for a large
number of qubits.  To do so, we need to be able to randomly choose an
element of $A$.

\begin{definition}
Consider the following coordinate map on the commutative group $A$:
\begin{equation}
[0,2\pi]^{N-1}\rightarrow A \mbox{ by }
(t_0,t_2,\cdots,t_{N-2}) \mapsto  
\mbox{exp } E_0 \big(
\sum_{j=0}^{N-2} i t_{j} \ket{j}\bra{j}-i t_{j}\ket{j+1}\bra{j+1}\big)E_0^\dagger
\end{equation}
The Haar measure on $da$ is the group multiplication invariant measure
$da= (2\pi)^{-N+1}dt_0\; dt_2 \; \cdots \; dt_{N-2}$.  This is the pushforward
of the independent product of uniform measures $dt_j/(2 \pi)$ on
each $[0,2\pi]$.
\end{definition}

Recall that for $p=2n$, Theorem \ref{thm:almost_all} asserts that
according to $da$, almost all $a \in A$ have $\kappa(a)=1$ for $p$ large.
Specifically, we assert
\begin{equation}
\mbox{lim}_{p \to \infty} da ( \{ a \in A \; ; \; \kappa(a)=1 \}) = 1
\end{equation}
We prove this assertion shortly, but we first need a lemma.

\begin{lemma}  
\label{lem:prob}
Label as uniform distribution on the circle a distribution
whose pullback to $[0,2\pi]$
under $t \mapsto \mbox{e}^{2\pi it}$ is uniform, and
similarly say two random variables $Z_1$, $Z_2$ on $\{ z \bar{z}=1\}$
are independent iff their pullbacks to $[0,2\pi] \times [0,2\pi]$ are.
Then suppose $Z_1$ is any random variable on the circle,
and let $Z_2$ be independent to $Z_1$ and uniform.  Then
$Z_1 Z_2$ is uniform.
\end{lemma}

\begin{proof}
Consider the random variable $T = -i \log Z_1 - i \log Z_2 \mbox{mod } 2\pi$ 
on $[0,2\pi]$.  Let $f_1(t)$ be the pullback probability density
function of the nonuniform random variable $Z_1$ to $[0,2\pi]$.
We let $F_T(t)= \mbox{Prob}(T \leq t)$ be the cumulative
density function.  Then 
\begin{equation}
F_T(t)=\frac{1}{2\pi}\int_0^{2\pi} \mbox{Prob}(-i \log Z_2 \in [s, s+t])
f_1(s) ds = \frac{1}{2\pi} \int_0^{2\pi} t f_1(s) ds = t/(2\pi)
\end{equation}
Since $F_T(t)=t/(2\pi)$, we see that $T$ is uniform.  Hence
$Z_1 Z_2$ is uniform.
\end{proof}

\begin{proof}[\ of Theorem \ref{thm:almost_all}]
First, let us check that $\{ \kappa(a)=1 \}$ is $da$-measurable.  To
see this, note that the concurrence spectrum $\lambda_c(a)$ may
be expressed in terms of the coordinates $t_j$ as follows:
\begin{equation}
d_0^2 = \mbox{e}^{2it_0}, d_1=\mbox{e}^{2it_1-2it_0},
d_2^2 = \mbox{e}^{2it_2-2it_1}, \cdots
d_j^2 = \mbox{e}^{2it_j-2it_{j-1}}, \cdots,
d_{N-1}^2 =\mbox{e}^{-2it_{N-2}}
\end{equation}
Thus Lemma \ref{lem:convex_hull} induces a measurable
condition on the $t_j$.

Continuing the proof, by direct calculation $Z^2$ is a uniform random
variable on the circle $\{z \bar{z}=1\}$ given that $Z$ is such.
Thus note that $d_0^2$, $d_2^2$, $d_4^2$, $\cdots$, $d_{N-1}^2$ are $p=N/2$
independent, uniform random variables by
Lemma \ref{lem:prob}.  It suffices to show that $\ell+1=p$ independent,
random variables on the circle have $0$ in their convex hull as $\ell \mapsto
\infty$.  Relabel $d_0^2=Z_0$, $d_2^2=Z_1$, $\cdots$ $d_{N-1}^2=Z_{\ell}$.

Without loss of generality, say $Z_0=1$.  Let $C_2$ be the event that
no $Z_1$, $Z_2$, $\cdots$ $Z_\ell$ is in the second
quadrant $\{ z=x+iy \; ; \; x<0, y>0 \}$, with $C_3$ similar for the
third quadrant $\{ x<0, y<0\}$.  
Let $D$ be the event that $0$ is in the convex hull
of $Z_0$, $Z_1$, $\cdots$, $Z_\ell$.  Then 
$(\mbox{NOT }C_2 \cap \mbox{NOT }C_3) \subset D$.  Then
$\mbox{Prob}(\mbox{NOT }C_2 \cap \mbox{NOT }C_3) \leq \mbox{Prob}(D)$, and
\begin{equation}
1 - \mbox{Prob}(D) \leq 1-\mbox{Prob}(\mbox{NOT }C_2 \ \mbox{ and NOT }C_3) =
\mbox{Prob}(C_2 \mbox{ or }C_3) = (1/2)^\ell
\end{equation}
Hence as $\ell \to \infty$, $\mbox{Prob}(D)$ goes to $1$.  Hence
the probability ${\tt CH}[\lambda_c(v)]$ contains $0$ limits to $1$.
\end{proof}

\section{Conclusions and Ongoing Work}

We have shown that there exists a generalized
canonical decomposition of unitary
operators on {\bf $n$} qubits which may be used to study changes in the
concurrence entanglement monotone.  This decomposition closely resembles
the older two-qubit decomposition when $n$ is even, and it may be
used to study the concurrence-entanglement capacity of generic unitary
operators.  The main result is that such a generic unitary
operator is almost always perfectly 
entangling with respect to the concurrence monotone when
the number of qubits is large and even.

Ongoing work would attempt to extend the dynamical viewpoint taken
in this paper.  Specifically, the unitary operator describes the
dynamics of a quantum data state, and the present techniques allow
us to quantitatively study the dynamics of the concurrence entanglement
measure.  Similarly, we would wish to study the dynamics of this 
concurrence capacity of quantum computations in naturally
defined families or sequences of such computations.  As a separate
topic, we might also study the failure of the concurrence function itself
by quantifying how entangled a quantum state with zero concurrence may
be.

\appendix
\section{Computing the {\tt CCD} When
the Number of Qubits Is Even}
\label{app:compute}

This appendix recalls how to compute the canonical decomposition
in an even number $n=2p$ of qubits.  Note that
other arguments in the case $n=2$
\cite{Khaneja:01, Lewenstein:01} 
may be found in the literature,
and that the present treatment is a straightforward genearlization of
a matrix-oriented treament in the two-qubit case
\cite[App.A]{Bullock:03}.  It is included for completeness.

The overall structure of the algorithm is contains two steps.
\begin{enumerate}
\item
\label{step:uSVD}  Produce an algorithm for computing the
decomposition $SU(N)= SO(N) \; D \; SO(N)$ for $D$ the diagonal
subgroup of $SU(N)$.  We will refer to this decomposition as
the \emph{unitary {\tt SVD} decomposition} henceforth.
\item  
\label{step:change_basis}
Recall $E_0$ the standard entangler of Definition \ref{def:stdE}.
Given a $v \in SU(N)$ for which we wish to compute
the {\tt CCD}, compute first the unitary
{\tt SVD} $E_0^\dagger v E_0=o_1 d o_2$.  Then we have a {\tt CCD} given by
\begin{equation}
v \; = \; (E_0 o_1 E_0^\dagger) (E_0 d E_0^\dagger) (E_0 o_2 E_0^\dagger)\; = \; k_1 a k_2
\end{equation}
since $k_1=E_0 o_1 E_0^\dagger \in K$, $k_2=E_0 o_2 E_0^\dagger \in K$, and
$a=E_0 d E_0^\dagger \in A$.
\end{enumerate}
Note that the unitary {\tt SVD} decomposition exists due to $KAK$
metadecomposition theorem, taking as inputs 
$G=SU(N)$, $\theta_{{\bf AI}}(X)=\bar{X}$, and $\mathfrak{a}$ the
diagonal subalgebra of $\mathfrak{su}(N)$.

Before continuing to Step \ref{step:uSVD}, we first prove a lemma.
It is useful in computing particular instances of the unitary {\tt SVD}.

\begin{lemma}
\label{lem:diag}
For any $p \in SU(N)$ with $p=p^T$,
there is some $o \in SO(N)$ such that $p=o d o^T$
with $d$ a diagonal, determinant one matrix.
\end{lemma}

\begin{proof}
We first show the following.
\begin{quotation}
$\forall\ a, b$, symmetric real $N \times N$ matrices with
$ab = ba$, there is some $o \in SO(N)$ such that $oao^T$ 
and $obo^T$ are diagonal.
\end{quotation}
It suffices to construct a basis which is simultaneously a basis of
eigenvectors for both $a$ and $b$.  Thus, say $V_\lambda$ is the $\lambda$
eigenspace of $b$.  
For $x \in V_\lambda$, $b(ax) = a (bx) = \lambda ax$, i.e.
$x \mapsto ax$ preserves the eigenspace.  Now find eigenvectors for $a$
restricted to $V_\lambda$, which remains symmetric.  Thus we may find
the desired $o \in SO(N)$, making choices of orderings and signs on
an eigenbasis as appropriate for determinant one.

Given the above, write $p=a+ib$. Now 
$ {\bf 1} = p p^\dagger = p \bar{p} = (a+ib)(a-ib) = (a^2 + b^2) + i(ba-ab)$.
Since the imaginary part of ${\bf 1}$ is ${\bf 0}$, we conclude that $ab=ba$.
Hence a single $o$ exists per the last paragraph which diagonalizes the
real and imaginary parts.
\end{proof}

Suppose then that $v=o_1 d o_2$ is the unitary {\tt SVD} of some
$v \in SU(N)$.  For convenience, we also label $v=po_3$ the 
type {\bf AI} Cartan decomposition
\cite[thm1.1.iii,pg.252]{Helgason:01} 
\cite[thm6.31.c]{Knapp:96}.  This is a generalized polar
decomposition in which $p=p^T$, $k\in SO(N)$.  Note that it is
equivalent via Lemma \ref{lem:diag} to compute $v=pk$,
as the unitary {\tt SVD} follows by $v=(o_1 d o_1^T) k= o_1 d o_2$.
Continuing to the algorithm for Step \ref{step:uSVD},
\begin{itemize}
\item{
Compute $p^2$ as follows:
$p^2 = p p^T = p o_3 o_3^T p^T=vv^T$.}
\item{
Apply Lemma \ref{lem:diag} to $p^2$.
Thus $p^2 = o_1 d^2 o_1^T$ for $o_1 \in SO(N)$.  
}
\item{
Choose square roots entrywise in $d^2$ to form
$d$.  Be careful to ensure $\mbox{det } d =1$.
}
\item{Compute $p = o_1 {d} o_1^T$.}
\item{Thus $o_3 = p^\dagger v$, and $v=p (o_3)=o_1 d o_1^T o_3=
o_1 d o_2$.}
\end{itemize}
This concludes the algorithm for computing the unitary
{\tt SVD} of Step \ref{step:uSVD}.

Step \ref{step:change_basis} is almost follows given the
inline description.  The reader may produce algorithms outputting
$E_0$.

Another question is computational efficiency.  This is ongoing
work, but we note immediately that an implementation
of the
spectral theorem of Lemma \ref{lem:diag} is required.  This will be
difficult with current technologies in $16+$ qubits.  Moreover,
in the range of $50$ to $60$ qubits 
an even spread of the concurrence spectrum $\lambda_c(v)$ 
of Definition \ref{def:con_spec} would make
certain elements indistinguishable at $16$-digit precision.

\section{Concurrence level sets and $K$ orbits}

Mathematically, related measures are often easier to use than
$C_n$.  For example, the concurrence quadratic form
$Q_n^C (\ket{\psi}) = \mathcal{C}_n (\ket{\psi},\ket{\psi})$
with $C_n(\ket{\psi}) = |Q_n^C(\ket{\psi})|$ has smaller
level sets than $C_n$ itself.  Moreover, it turns out 
that the normalized states within these
level sets $[Q^C_n]^{-1}(\{z\}) = 
\{ \ket{\psi} \; ; \; Q^C_n (\ket{\psi})=z\}$
are naturally orbits of the group $K$, which must then
be false for $C_n$.

Suppose throughout $n=2p$ is an even number of qubits.
For a vector $v \in \mathbb{C}^N$,
put $Q_{{\bf AI}} (v) = v^T v$, noting that $Q_{\mathcal{C}} (E_0 v)=
Q_{\bf AI} (v)$.  Moreover, for $O \in SO(N)$, we have the following:
\begin{equation}
\label{eq:translate}
Q_{\bf AI}(O \cdot v) = Q_{\mathcal{C}}[E_0 O E_0^\dagger \cdot (E_0 v)]
\end{equation}
Thus we may study level sets of $Q_{\bf AI}$ under $SO(N)$ rather
than study level sets of $Q_{\mathcal{C}}$ under $K$.  Now if
$v=v_1 + i v_2$ is a decomposition into real and imaginary parts of
a complex vector, note that
$Q_{\bf AI} (v_1+iv_2) = v^T v = (|v_1|^2 -|v_2|^2)+2i(v_1\cdot v_2)$.

\vbox{
\begin{lemma}
Label $S^{2N-1}=\{ \ket{\psi} \; ; \; \langle \psi \ket{\psi} = 1\}$.
We have the following orbit decompositions of the
level sets $Q_{\bf AI}^{-1}(\alpha) \cap S^{2N-1}$ for any
fixed $\alpha \in \mathbb{C}$.
\begin{enumerate}
\item Let $t$ be real, and let $v \in Q_{\bf AI}^{-1}(t) \cap S^{2N-1}$.
Then $Q_{\bf AI}^{-1}(t) \cap S^{2N-1} = [SO(N) \cdot v]$.
\item  Let $\alpha$ be complex, and let 
$v \in Q_{\bf AI}^{-1}(\alpha) \cap S^{2N-1}$.  Then
$Q_{\bf AI}^{-1}(\alpha) \cap S^{2N-1} = [SO(N) \cdot v]$.
\end{enumerate}
\end{lemma}
}

\begin{proof}
For the first item, write $v=v_1+iv_2$.  Then $v_1 \cdot v_2$ is zero
as a set of real vectors.  Consider the subset of
$\mathbb{R}^{2N}$ given by $|v_1|^2-|v_2|^2=t$.  Suppose now
we have another pair of orthogonal vectors
$w_1$, $w_2$ with $|w_1|^2 - |w_2|^2=t$ and
$|w_1|^2+|w_2|^2=1$.  Then $|w_1|^2 = |v_1|^2 = (1-t)/2$,
thus $|v_2|^2=|w_2|^2$ so that there is some $O \in SO(N)$
with $O \cdot v_1 = w_1$, $O \cdot v_2=w_2$.

For the second item, suppose $\alpha = \mbox{e}^{i \phi} t$ for
some $t \in \mathbb{R}$.  Now if $v \in Q_{\bf AI}^{-1}(\alpha)$,
then note that we have
$Q_{\bf AI}(\mbox{e}^{-i\phi/2}v)=\mbox{e}^{-i\phi}Q_{\bf AI}(v)
= \mbox{e}^{-i\phi} \alpha = t$.  Conversely, if $w \in Q^{-1}_{\bf AI}(t)$
we have $\mbox{e}^{i\phi/2} w \in Q^{-1}_{\bf AI}(\alpha)$.  Having
established bijective phase maps between the two level sets,
it must also be the case that the level set of $\alpha$ forms a single
$SO(N)$ orbit.
\end{proof}

\begin{corollary}
The restricted action of $K$ to the normalized kets in any
concurrence level set is transitive.  Specifically, suppose
$\alpha \in \mathbb{C}$, with $\ket{\psi}$ normalized with
$Q_{\mathcal{C}}(\ket{\psi})=\alpha$.  Then
label $S^{2N-1}=\{ \bra{\phi} \phi \rangle =1\}$ the set of normalized
kets.  Per Equation \ref{eq:translate}, we have
$K \cdot \ket{\psi} = Q^{-1}_{\mathcal{C}}(\alpha) \cap S^{2N-1}$.
\end{corollary}

We restate the result colloquailly.  Should any two normalized
states $\ket{\phi}$, $\ket{\psi}$ have the same concurrence, then
there is some global phase $\mbox{e}^{i\theta}$ so that
$\ket{\phi} = \mbox{e}^{i\theta} k \ket{\psi}$ for $k \in K=
E_0 \; SO(N) \; E^\dagger_0$.

\section{Concurrence is an entanglement monotone}
\label{app:monotone}

The $n-$tangle, defined to be $\tau_{n}(|\psi\rangle)=C_{n}(|\psi\rangle)^{2}$ 
has been proposed \cite{Wong:01} as a measure of $n$ qubit 
entanglement for $n$ even.  The $n$-tangle of a state $|\psi\rangle$, like 
the $n$-concurrence, assumes real values in the range 
$0\leq \tau_{n}\leq1$ and has been shown to be an entanglement monotone, 
meaning $\tau_{n}$ is a convex function on states and is non-increasing under 
local operations and classical communication (LOCC).  Most of our arguments focus on constructions more directly
related to the concurrence $C_n$ rather than the $n$-tangle
$\tau_n=(C_n)^2$.   Therefore, for completeness, we show that 
the $n$-concurrence is, in fact, a good measure of entanglement.  The 
monotonicity property of a function is established by considering 
its action on mixtures of quantum states encoded within Hermitian
density matrices $\rho$ with $\mbox{tr } \rho=1$.  
See, e.g., \cite{Nielsen:00}.

\begin{appendix_definition}
\label{def:conc_of_density}
The $n$-concurrence can be defined on mixed states 
$\rho$ using the convex roof extension:
\begin{equation}
C_{n}(\rho)=\min
\Bigg\{ \sum_{k}\lambda_{k}C_{n}(|\psi^{k}\rangle) \; ; \; \quad
\rho = \sum_{k} \lambda_{k} \ket{\psi^k} \bra{\psi^k}, \ \ 
\ket{\psi}^k \in \mathcal{H}_n, 
\ \ \bra{\psi^k} \psi^k \rangle = 1 \Bigg\}
\end{equation}
This minimization is over all pure state ensemble 
decompositions of the state 
$\rho=\sum_{k}\lambda_{k}|\psi^k\rangle\langle \psi^k|$.
\end{appendix_definition}

This definition is quite intricate.  We point out the following
remarkable result, not used in the sequel.

\begin{theorem}[Uhlmann,\ \ \cite{Uhlmann:99,Wong:01}]
We may express $C_n(\rho)$ in closed form as follows:
\begin{equation}
C_{n}(\rho)=\max\{0,\lambda_{0}-\lambda_{1}\ldots-\lambda_{N-1}\} \;
\end{equation}
Here, the $\lambda_k$ are the square roots of 
the eigenvalues (in non-increasing order) of the 
product $\rho \tilde{\rho}$ where $\tilde{\rho}=S\overline{\rho}S^{-1}$.  
\end{theorem}

The necessary and sufficient conditions for a function on quantum states 
to be a entanglement monotone are delineated in \cite{Vidal:00}.  For 
the $n$-concurrence, they can be summarized as follows:
\begin{itemize}
\item $C_n\geq 0$, and $C_n(\rho)=0$ if $\rho$ is fully separable.
\item $C_n$ is a convex function, i.e. 
$C_n(p \rho_1+(1-p) \rho_2)
\leq p C_n(\rho_1)+(1-p) C_n(\rho_2), \forall p\in[0,1]$ 
and $\rho_1,\rho_2$ Hermitian matrices of trace one
\item $C_n$ is non increasing under LOCC.  Specifically, 
$C_n(\rho)\geq \sum_j p_j C_n(\rho_j)$, where 
$\rho_j=A_j\rho A_j^\dagger/p_j$ are the states conditioned on the 
outcome $j$ of a positive operator valued measurement 
({\tt POVM}) which occurs with probability $p_j=tr[A_j^\dagger A_j\rho]$.
\end{itemize}

Before proving this, we first establish the useful fact that the 
$n$-concurrence is invariant under permutations of the 
qubits.  Defining $\Pi_{n}$ to be the set of unitary operators 
corresponding to permutations on $n$ qubits, we have:
\begin{proposition}  For $n$ even, $C_{n}(P|\psi\rangle)=C_{n}(|\psi\rangle)\  
\forall P\in \Pi_{n}$.
\label{prop:permsinK}
\end{proposition}

\begin{proof}
Any permutation $P$ on $n$ elements can be written as a finite composition 
of transpositions on pairs of elements.
Hence it suffices to show invariance 
under a single swap operation.  Writing the swap operator between qubits 
$j$ and $k$ as 
\begin{equation}
S_{jk}=\frac{{\bf 1}_{j}\otimes{\bf 1}_{k}+\sigma_{j}^{x}\otimes\sigma_{k}^{x}+
\sigma_{j}^{y}\otimes\sigma_{k}^{y}+ 
\sigma_{j}^{z}\otimes\sigma_{k}^{z}}{2},\;
\end{equation}
we have for any state $|\psi\rangle$,
\begin{equation}
\begin{array}{lclcl}
C_{n}(S_{jk}|\psi\rangle)&=&|\overline{\langle 
\psi| S_{jk}^\dagger}SS_{jk}|\psi\rangle|
&=&|\overline{\langle\psi|}S_{jk}SS_{jk}|\psi\rangle|\\
&=&|\overline{\langle\psi|}SS_{jk}^{2}|\psi\rangle|
&=&C_{n}(|\psi\rangle). \\
\end{array}
\end{equation}
Here we have used the fact that $S_{jk}$ is real symmetric and unitary, and 
in the third equality we use the fact that 
$[\sigma^{y}_{j}\otimes\sigma^{y}_{k},\sigma^{l}_{j}\otimes\sigma^{l}_{k}]=0$ 
for $\sigma^{l}\in\{\sigma^{x},\sigma^{y},\sigma^{z}\}$.  This proposition 
necessarily implies that $\Pi_{n}^{+}\subsetneq K$, where $\Pi_{n}^{+}$ is 
the set of unitary permutation matrices on $n$ objects with $+1$ 
determinant, i.e. permutations composed of an even number of transpositions.
\end{proof}

\begin{lemma}  $C_{n}(\rho)$ is an entanglement monotone.
\label{monotone}
\end{lemma}
\begin{sketch}
For the first condition, one first checks that
$0\leq C_n\leq 1$ using the eigenvalue decomposition of the matrix
$S=(-i\sigma^y_1)(-i\sigma^y_2)\cdots(-i\sigma^y_n)$.  Then any 
separable state can be realized by 
stochastic local unitaries acting on the fiducial separable state 
$|0\rangle _n=|0_1\ldots0_n\rangle$.   Now, 
$C_n(|0\rangle _n)=0$ and $C_n$ is invariant under local unitaries
per Proposition \ref{prop:localsinK} and Theorem \ref{thm:K_sym}.
To generalize from pure states to density matrices, 
recall Definition \ref{def:conc_of_density}. 

The second condition is shown by writing the minimal ensemble decompositions for $\rho_1$ and $\rho_2$ separately as
\begin{equation}
p \min_{\{\lambda_k,|\psi^k\rangle\}|\sum{\lambda_k|\psi^k\rangle\langle \psi^k|=\rho_1}}\sum_{k}\lambda_k C_n(|\psi^k\rangle)+(1-p) \min_{\{\beta_k,|\phi^k\rangle\}|\sum{\beta_k|\phi^k\rangle\langle \phi^k|=\rho_2}}\sum_{k}\beta_k C_n(|\phi^k\rangle).\;
\end{equation}
These are not necessarily the minimal decompositions for the composite state $\rho=p p_1+(1-p) p_2$, therefore, $C_n(p \rho_1+(1-p) \rho_2)\leq p C_n(\rho_1)+(1-p) C_n(\rho_2)$. 

Finally, we show that the $n$-concurrence 
is on average non-increasing under LOCC.  First, because of permutation 
symmetry 
of the concurrence we can consider operations on one particular qubit of 
the $n$ qubit system, say the first.   An arbitrary, trace perserving, 
completely positive map on a quantum system can written in the Krauss 
decompostion \cite{Krauss:83} as $S(\rho)=\sum_{j}A_{j}\rho A_{j}^\dagger$ 
where the positive Krauss operators satisfy the sum rule 
$\sum_{j}A_{j}^\dagger A_{j}={\bf 1}$.  The map can be composed of multiple 
operations with two operators at a time so we consider only two 
operators $A_{0}$ and $A_{1}$ acting on the first qubit.  By the polar 
decomposition theorem, the operators can be written as $A_{j}=u_{j}b_j$, 
where $b_j=\sqrt{A^\dagger_jA_j}$ is positive 
and $u_j$ is defined to be ${\bf 1}$ on the 
kernel $\mathcal{K}$ of $A_j$ and $A_j|A_j|^{-1}$ 
on $\mathcal{K}^{\perp}$.  Physically, the map $S(\rho)$ corresponds 
to a generalized measurement on $\rho$ followed by 
a unitary operation conditioned on the measurement.  Because 
of the sum rule, which corresponds to trace preservation, we can 
write $A_0=w_o\cos{gX}$ and $A_1=w_1\sin{gX}$ for 
$g\in {\mathbb{R}}$ and $X$ a positive operator with unit trace.  These 
operators are expressed in simpler form by diagonalizing 
$X$, viz. $A_0=u_od_ov$ and $A_1=u_1d_1v$, where 
$u_{j},v\in SU(2)$ and $d_{j}$ are real diagonal matrices with 
elements $(q,r)$ and $(\sqrt{1-q^2},\sqrt{1-r^2})$.  The average 
concurrence of a state $\rho$ after the 2 outcome POVM is
\begin{equation}
\begin{array}{lll}
p_0C_n(\rho_0)+p_1C_n(\rho_1)&=&p_0  
\min_{\{\lambda_k,|\psi^k\rangle\}|
\sum{\lambda_k|\psi^k\rangle\langle \psi^k|=\rho}}\sum_{k}\lambda_kC_n(A_0|\psi^k\rangle/\sqrt{p_0})\\
& &+p_1 \min_{\{\beta_k,|\phi^k\rangle\}|\sum{\beta_k|\phi^k\rangle\langle \phi^k|=\rho}}\sum_{k}\beta_k C_n(A_1|\phi^k\rangle/\sqrt{p_1}).\;
\end{array}
\end{equation}
States conditioned on the first outcome satisfy:
\begin{equation}
\begin{array}{lll}
C_{n}(A_0|\psi\rangle/\sqrt{p_0})&=&|\overline{\langle \psi|}v^Td_{0}^Tu_{0}^TSu_0d_{0}v|\psi\rangle|/p_{0}\\
&=&qrC_{n}(v|\psi\rangle)/p_{0}\\
&=&qrC_{n}(|\psi\rangle)/p_{0},\;
\end{array}
\end{equation}
where in the second equality we have used 
$u_0^T(-i\sigma^{y})u_0=-i\sigma^{y}$, and the third equality follows by invariance of concurrence under local unitaries.  Similarly, $C_{n}(A_1|\phi\rangle/\sqrt{p_1})=\sqrt{(1-q^2)(1-r^2)}C_{n}(|\phi\rangle)/p_{1}$.
The result is,
\begin{equation}
\begin{array}{lll}
p_0C_n(\rho_0)+p_1C_n(\rho_1)&=&qr  \min_{\{\lambda_k,|\psi^k\rangle\}|\sum{\lambda_k|\psi^k\rangle\langle \psi^k|=\rho}}\sum_{k}\lambda_k C_n(|\psi^k\rangle)\\
& &+\sqrt{(1-q^2)(1-r^2)} \min_{\{\beta_k,|\phi^k\rangle\}|\sum{\beta_k|\phi^k\rangle\langle \phi^k|=\rho}}\sum_{k}\beta_k C_n(|\phi^k\rangle)\\
&=&(qr+\sqrt{(1-q^{2})(1-r^{2})})C_{n}(\rho)\\
&\leq&C_n(\rho),\;
\end{array}
\end{equation}
with equality iff $q=r$, i.e. the $A_{j}$ are stochastic unitaries.
\end{sketch}

\end{document}